\documentclass{aa} 

\usepackage[varg]{txfonts}
\usepackage[T1]{fontenc}
\usepackage[utf8]{inputenc}
\usepackage{amssymb}
\usepackage[]{hyperref}
\usepackage{amsmath}
\usepackage{graphicx}
\usepackage[dvipsnames]{xcolor} 
\usepackage[normalem]{ulem}

\def\aap{A\&A}

\def\apjs{ApJSS}

\begin{document}
		
\title{Computational general relativistic force-free electrodynamics: II. Characterization of numerical diffusivity}
\titlerunning{Computational GRFFE: Characterization of numerical diffusivity}

\author{J. F. Mahlmann\inst{1}\thanks{jens.mahlmann@uv.es}, M. A. Aloy\inst{1}, V. Mewes\inst{2,3},P. Cerdá-Durán\inst{1}}
\institute{Departament d'Astronomia i Astrofísica, Universitat de València, 46100, Burjassot (Valencia), Spain
	\and
	National Center for Computational Sciences, Oak Ridge National Laboratory, P.O. Box 2008, Oak Ridge, TN 37831-6164, USA
	\and
	Physics Division, Oak Ridge National Laboratory, P.O. Box 2008, Oak Ridge, TN 37831-6354, USA}

\date{Received \textit{Month Day, Year}; accepted \textit{Month Day, Year}}

\abstract{Scientific codes are an indispensable link between theory and experiment; in (astro-)plasma physics, such numerical tools are one window into the universe's most extreme flows of energy. The discretization of Maxwell's equations - needed to make highly magnetized (astro)physical plasma amenable to its numerical modeling - introduces numerical diffusion. It acts as a source of dissipation independent of the system's physical constituents. Understanding the numerical diffusion of scientific codes is the key to classifying their reliability. It gives specific limits in which the results of numerical experiments are physical. We aim at quantifying and characterizing the numerical diffusion properties of our recently developed numerical tool for the simulation of general relativistic force-free electrodynamics by calibrating and comparing it with other strategies found in the literature. Our code correctly models smooth waves of highly magnetized plasma. We evaluate the limits of general relativistic force-free electrodynamics in the context of current sheets and tearing mode instabilities. We identify that the current parallel to the magnetic field ($\mathbf{j}_\parallel$), in combination with the breakdown of general relativistic force-free electrodynamics across current sheets, impairs the physical modeling of resistive instabilities. We find that at least eight numerical cells per characteristic size of interest (e.g., the wavelength in plasma waves or the transverse width of a current sheet) are needed to find consistency between resistivity of numerical and of physical origins. High-order discretization of the force-free current allows us to provide almost ideal orders of convergence for (smooth) plasma wave dynamics. The physical modeling of resistive layers requires suitable current prescriptions or a sub-grid modeling for the evolution of $\mathbf{j}_\parallel$.}

\maketitle

\section{Introduction}
\label{sec:Introduction}

The numerical modeling of the dissipation, transport, and emission of  high-energy particles 
by strongly magnetized plasma is a necessary ingredient in the theoretical interpretation of 
the highly energetic astronomical phenomena associated with compact objects such as neutron stars 
(NSs) and black holes (BHs). 
Current sheets seem to be one important location where such processes take place. In these, reconnection of the magnetic field results in the acceleration of relativistic particles 
\citep[see, e.g.,][for some recent examples]{Ball2019,Petropoulou2019,Kilian2020}, as well as the locations of Fermi-type processes \citep{Guo2019}. 

When the magnetic diffusivity (or resistivity, $\eta$) is sufficiently large, plasma dynamics 
change due to nonideal processes and the plasma can no longer be modeled as ideal, that is, as a perfectly conducting plasma. 
Astrophysical plasma is typically an environment of extremely low magnetic diffusivity. Thus, resorting to ideal (relativistic) magnetohydrodynamics (MHD) or force-free electrodynamics 
(FFE) is a reliable assumption when magnetic fields dominate all plasma dynamics. However, a 
numerical treatment of the challenges at hand requires introducing a controlled 
(often implicit) amount of numerical diffusivity. Such a diffusivity of numerical 
origin stems from two main sources: first, from the discretization of a set of physical 
(balance) laws, often partial differential equations; and second, from the need to stabilize 
numerical solutions across the various types of discontinuities that exist in ideal MHD or FFE.

The design of numerical codes commonly focuses on minimizing the amount of numerical diffusivity 
across discontinuities. Howewer, much less weight is placed on minimizing or (at least) 
characterizing the numerical diffusivity resulting from the discretization of the governing 
equations (with considerable exceptions, e.g., \citealt{Rembiasz2017}, in Eulerian MHD, or \citealt{Obergaulinger_2020arXiv200101927}, in Newtonian hydrodynamics). We 
find that a thorough account of the numerical diffusivity of algorithms designed to solve 
the equations of FFE or general relativistic FFE (GRFFE) is of the utmost importance. We therefore 
assess whether numerical diffusivity behaves as a physical diffusivity and if it introduces 
pathological biases in the modeling of (astrophysical) plasma. This is a necessary step in grading the quality of numerical results when the FFE approximation reaches its limit. Such 
limits are specifically expected at the location of current sheets, (smooth) Alfv\'{e}n waves, 
and Alfv\'{e}n discontinuities.

The linear phase of the dynamics in pinched force-free current sheets has been modeled in 
FFE \citep{Komissarov_etal_2007MNRAS.374..415} and compared to particle-in-cell simulations 
\citep{Lyutikov_2018JPlPh..84b6301}. Still, conventional FFE codes are insufficient to account 
for the nonideal electric fields that drive complex dynamics. In contrast to the limits 
of nonvanishing particle inertia (MHD), energy is dissipated by violations of the algebraic 
constraints under which the FFE approximation is valid. In this second part of our two-series 
publication describing our new GRFFE code, we aim to quantify the impact of such 
nonideal dissipation in both smooth plasma waves and current sheets as thin current-carrying 
layers across which the magnetic field either changes direction or changes in magnitude \citep{Harra2004}. 
Aiming from a comprehensive numerical modeling of (astrophysical) plasma across magnetization 
regimes, we ask the questions of why FFE methods
fail to resolve the dynamics in current-dominated domains, and what the physical consequences of these failures are.

This second manuscript in our series of publications is dedicated to answering these questions, 
evaluating previous results obtained by our code, and providing leverage points at which 
kinetic plasma modeling may bridge the limits of FFE. In the context of instabilities in 
magnetar magnetospheres, \citet{Parfrey2013}, \citet{Carrasco_etal_2019_10.1093/mnrasl/slz016}, and \citet{Mahlmann2019} 
have encountered current sheets at the stellar surface, and sensitivity to a (conservative) 
transport of charges throughout the domain. The longest variability timescales in the observed TeV 
emission, for example, in M87, have recently been linked to recurring periods of efficient 
Blandford/Znajek \citep{Blandford1977} type outflows induced by the accretion of magnetic 
flux tubes \citep{Parfrey2015,Mahlmann2020}. Reconnection and plasmoid formation in BH accretion processes are likely to act on much shorter timescales \citep[studied in the ideal limit by, e.g.,][]{nathanail2020} and involve relevant physical nonideal electric fields \citep[analyzed in the resistive limit by, e.g.,][]{Ripperda2020}. A large array of work makes use of numerical laboratories set up in (GR)FFE in order to simulate the 
most extreme environments of the universe while constantly breaking their own limits. 
Understanding the pathology of such breaches, and providing valid test cases for their evaluation, 
has motivated this additional technical exploration of our recent code development effort.

This work is organized as a series of papers. This manuscript (Paper II) characterizes the 
numerical resistivity of our GRFFE code \citep[as introduced in][Paper I]{Mahlmann2020b} in 
depth by studying the 1D diffusion of plasma waves (Sect.~\ref{sec:PlasmaWaves}) and the growth of 
2D tearing modes (TMs; Sect.~\ref{sec:TearingModes}) under force-free conditions. 
In Sect.~\ref{sec:Beyond_Ideal}, we explore a phenomenological current, effectively allowing 
for modeling beyond ideal FFE, namely, driving the evolution with a physical resistivity $\eta$.
We discuss the implications of our results on GRFFE methods in Sect.~\ref{sec:Discussion} 
and present our conclusions in Sect.~\ref{sec:conclusions}. We use units with $G=c=M_\odot=1$, as in Paper 1.

\section{Assessment of numerical resistivity}
\label{sec:Numerical_Resistivity}

In this section, we quantify the numerical resistivity of the GRFFE code presented 
in Paper I. This analysis is analogous to the one in \citet{Rembiasz2017} for Eulerian MHD codes, 
but applied to the case of FFE. For our analysis, we employ two key techniques: i) measuring 
damping rates of plasma waves that are captured in a 1D periodic domain for 
a large number of dynamical timescales (Sect.~\ref{sec:PlasmaWaves}); and ii) Measuring growth 
rates of TM perturbations in a 2D force-free current sheet 
(Sect.~\ref{sec:TearingModes}). A similar strategy was followed in \cite{Miranda-Aranguren2018} 
in resistive relativistic MHD. Since the development of TMs requires, at the minimum, a 
longitudinal current sheet in 2D (though, obviously, they can also develop in 3D), the study 
of the growth rate of resistive TMs allows us to quantify the numerical resistivity of 
our code in more than one dimension.

\begin{figure*}
	\centering
	\includegraphics[width=1.0\textwidth]{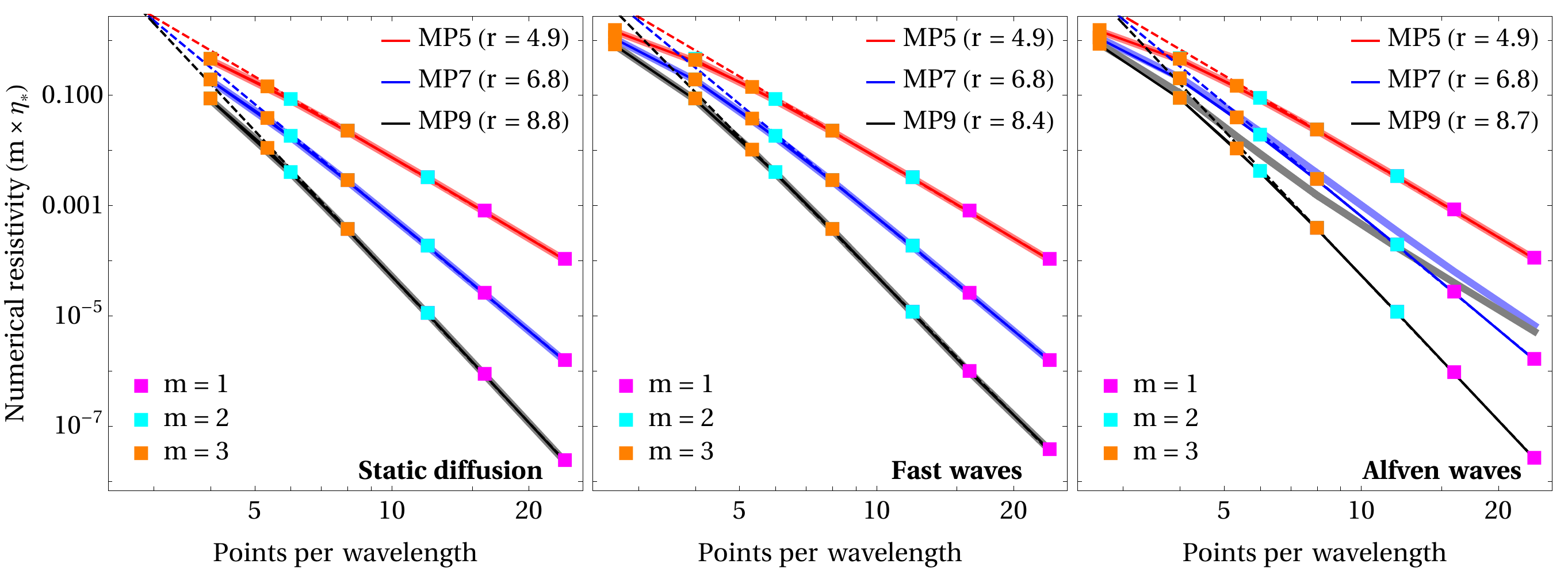}
	\caption{Assessment of numerical resistivity (normalized, see Eq.~\ref{eq:etafitnormalized}) 
		as a function of grid points per wavelength for a 1D static diffusion test (series A) 
		as well as propagating 1D plasma waves (series B and C, i.e., fast and Alfv\'{e}n). Numerical fit 
		parameters (Table~\ref{tab:resistivity_waves_table}) are obtained for the data points 
		of the $m=1$ mode (dashed lines). For series C, the discretization order of the numerical 
		derivatives in the calculation of 
		the current $\mathbf{j}_\parallel$ has significant impact on the order of convergence. 
		Thick lines in the background employ standard fourth-order finite differences in the calculation of
		reconstruction of $\mathbf{j}_\parallel$, while solid thin lines display the improved high-order 
		finite differences (Sect.~3.4, Paper I). }
	\label{fig:ResistivityWaves}
\end{figure*}
\begin{figure*}
	\centering
	\includegraphics[width=0.9\textwidth]{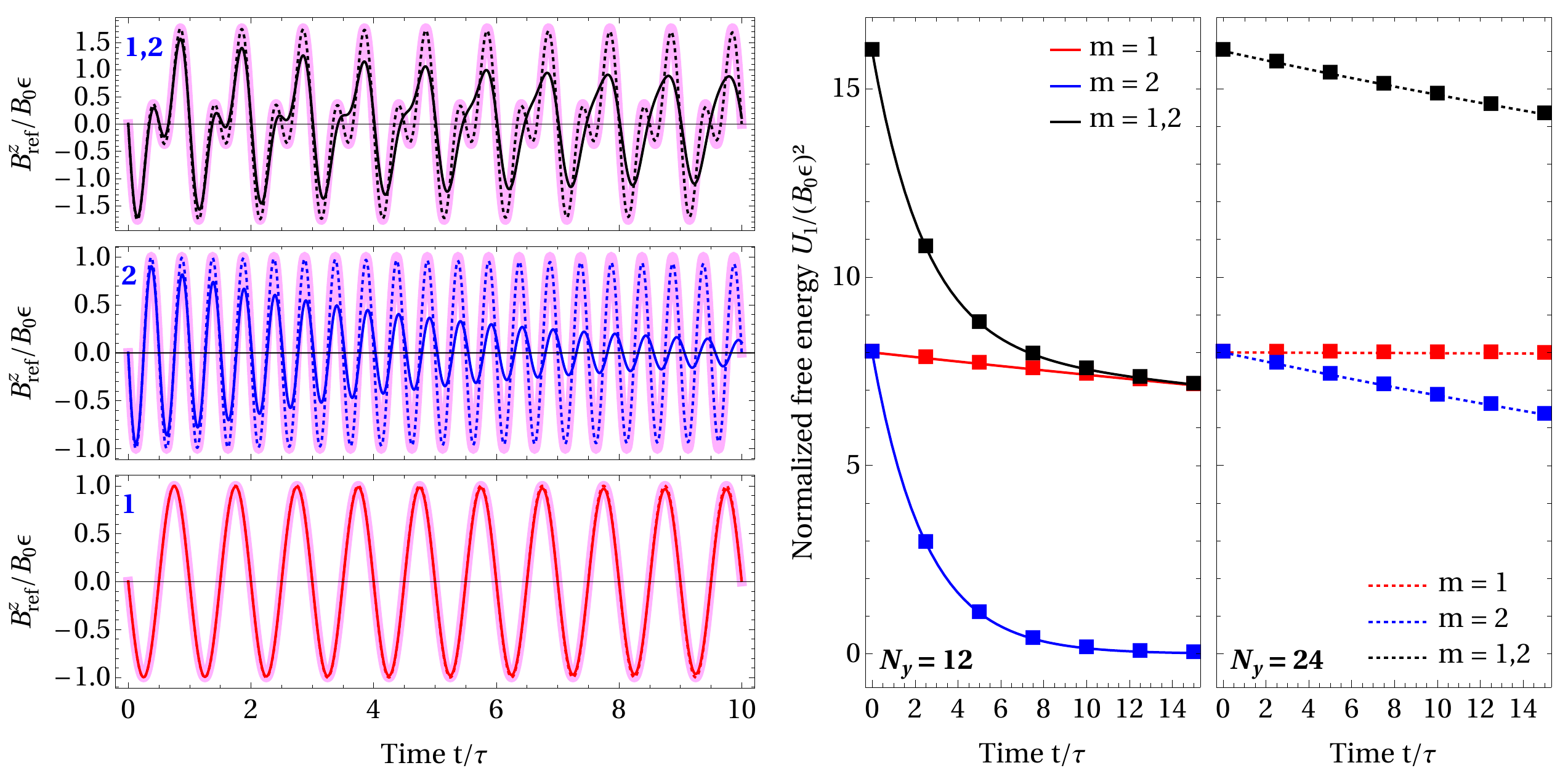}
	\caption{Exemplary amplitude evolution of selected modes of the fast wave model (showing $m=1$, $m=2$, 
		as well as the superposition $m=1,2$). The time is normalized to the light crossing time 
		of the computational domain ($\tau$). \textit{Left:} Field amplitude at the point $y=0$ 
		for a superposition of modes $m=1,\,2$ (top panel), $m=2$ (middle), and $m=1$ (bottom). 
		The analytic solution is indicated in the background (thick magenta lines). Different 
		resolutions are visualized by solid ($N_y=12$) and dotted ($N_y=24$) lines. \textit{Right:} 
		Evolution of the energy contained in the perturbations, given by Eq.~(\ref{eq:freeenergy}), 
		scaled to $\epsilon^2 B_0^2$ (squares) and theoretical model (solid lines for $N_y=12$ and dashed lines for $N_y=24$) employing the 
		fit parameters from Table~\ref{tab:resistivity_waves_table}. Mixed modes evolve according 
		to a superposition of two exponentially damped plasma waves with the appropriately scaled 
		damping rates inferred from Fig.~\ref{fig:ResistivityWaves}.}
	\label{fig:ResistivityOscillations}
\end{figure*}
%

\subsection{Diffusion of 1D plasma waves}
\label{sec:PlasmaWaves}

One of the simplest 1D assessments of numerical resistivity is checking the capacity of the code for maintaining standing or propagating waves over significant numbers 
of light-crossing times \citep[][]{Rembiasz2017}. Hereafter, quantities without any subscript 
refer to the total electromagnetic field, which we assume to be composed of a background field 
(annotated with subscript $0$) and a perturbation (annotated with subscript $1$), namely,
\begin{align}
	\mathbf{B}=\mathbf{B}_0+\mathbf{B}_1, \qquad \mathbf{D}=\mathbf{D}_0+\mathbf{D}_1, \qquad \rho = \rho_0 + \rho_1,
\end{align}
which correspond to the magnetic field, the electric field, and the charge density, respectively. 
These are the main evolved variables in our code, apart from the two potentials that we used to 
preserve the value of the divergence of the fields (which we always initialize to zero, see Paper I). 
We performed three series of 1D simulations for different kinds of waves with the goal of 
measuring the damping rate induced by the existence of a finite numerical diffusivity. 

Series A corresponds to a standing wave in FFE, which is excited by initializing the variables to
\begin{align}
	\mathbf{B}_0&=\mathbf{D}_0=\left(0,0,0\right), & \rho_0 &= 0, \nonumber \\
	\mathbf{B}_1&=\left(B_0\sin ky,0,B_0\cos ky\right), & \mathbf{D}_1&=\left(0,0,0\right), \label{eq:standingwavesetup}
\end{align}
where $k$ is the wavenumber. $\rho_1$ is initialized numerically to ensure $\nabla\cdot D_1 = \rho_1$ 
(here and in all the tests hereafter). $B_0$ is the scale-free amplitude of the standing wave, 
which we set to $B_0=1$ for the sake of simplicity. 

Series B corresponds to fast magnetosonic waves traveling in the direction of a guide field 
$\mathbf{B}_0$ along the $y$-direction, which are excited by the following initial perturbation 
\citep{Punsly2003}:
\begin{align}
	\mathbf{B}_0&=\left(0,B_0,0\right), & \mathbf{D}_0 &=\left(0,0,0\right), & \rho_0 &=0, \nonumber\\
	\mathbf{B}_1&=\left(B_0\epsilon\cos ky,0,0\right), & \mathbf{D}_1&=\left(0,0,B_0\epsilon\cos ky\right). && \label{eq:fastwavesetup}
\end{align}
Here, the amplitude of the perturbations is controlled by the dimensionless parameter  $\epsilon$. 
We chose the perturbation amplitude to be sufficiently small, $\epsilon=10^{-5}$, to guarantee 
that the perturbation can be regarded as linear with respect the FFE equations. This will 
allow us to measure the numerical resistivity from the damping rate of the waves (see below). 

Series C corresponds to Alfv\'{e}n waves traveling along the x-direction at an angle $\theta=\pi/4$ 
to the guide field. The initial perturbations in this case are \citep{Punsly2003}: 
\begin{align}
	\mathbf{B}_0&=(B_0/\sqrt{2},0,B_0/\sqrt{2}), & \mathbf{D}_0&=\left(0,0,0\right), \quad \rho_0 = 0,\nonumber\\ 
	\mathbf{B}_1&=\left(0,B_0\epsilon\cos kx,0\right), &
	\mathbf{D}_1&=\frac{B_0\epsilon}{\sqrt{2}}\left(\cos kx,0,-\cos kx\right).
	\label{eq:Alfvenwavesetup}
\end{align}
where $\epsilon$ is of the same value as in series B. In all series, the 1D domain is discretized 
in the $y$-direction with an extent $L=16$ in the domain $\left[-L/2,L/2\right]$. The initial 
data for the different cases is evolved for 125 light crossing times ($\tau=L$) of the domain. 
We impose periodic boundary conditions, so we chose a wavenumber $k=2\pi m/L$, with $m\in\mathbb{N}$.
Simulations of the cases $m=1, 2$ and $3$ (and linear combinations) are performed using different 
numerical resolutions to study the convergence of the method. We use the monotonicity preserving \citep[MP,][]{Suresh1997} 
reconstruction with three different orders: MP5, MP7 and MP9. 
In all cases we use a CFL factor of $f_\textsc{cfl}=0.2$.

For incompressible, viscous-resistive MHD, \cite{Campos1999PhPl....6...57} found that the 
damping rate, $\mathfrak{D}$, of Alfv\'{e}n and fast magnetosonic waves is proportional to the 
square of the wavenumber and to the diffusivity, $\xi$, namely
\begin{align}
	\mathfrak{D}=\frac{k^2}{2}\xi.
	\label{eq:Dampingrate}
\end{align}
Here, the diffusivity is a linear combination of the shear and bulk viscosity as well as the 
resistivity, $\eta$. Expression~(\ref{eq:Dampingrate}) strictly holds in the so-called weak 
damping approximation, namely, $k^4\xi^2/(4v_{\rm A}^2)\ll 1$ (where $v_{\rm A}$ is the 
Alfv\'{e}n velocity). 

\cite{Komissarov_etal_2007MNRAS.374..415} has shown that the equations of FFE in the so-called 
incompressible limit can be cast into a form nearly identical to that of incompressible MHD. 
In the incompressible limit, the drift speed $\mathbf{V}=\mathbf{D}\times\mathbf{B}/B^2$ 
fulfills $\nabla\cdot\mathbf{V}=0$. For models of the series A, $\mathbf{V}=0$ and 
$\nabla\cdot\mathbf{V}=0$ strictly hold, while for the series B and C, 
$\mathbf{V}=\mathcal{O}(\epsilon^2)\approx 0$ and $\nabla\cdot\mathbf{V}=\mathcal{O}(\epsilon^2)\approx 0$ 
to first order in $\epsilon$. Hence, for our choice of $\epsilon$, we can safely make the 
assumption that the damping of Alfv\'{e}n and fast modes in incompressible FFE proceeds 
analogously to their incompressible MHD counterparts.

Since FFE neglects the thermal and inertial contributions of the plasma, its equations are 
not affected by either bulk or shear viscosity (neither of physical nor of numerical origin). 
Thus, the diffusivity may only come from resistive effects. Provided that in FFE 
there is no physical resistivity, the diffusivity in our numerical results comes exclusively 
from the numerical resistivity, $\eta_*$. Following \cite{Campos1999PhPl....6...57} with the 
notation of \cite{Rembiasz2017}, we find $\xi=\eta$ for all three series of initial data,
\begin{align}
	\mathfrak{D}=\frac{k^2}{2}\eta\equiv\frac{k^2}{2}\eta_{*}\qquad\Longleftrightarrow\qquad\eta_{*}=\frac{2}{k^2}\mathfrak{D}\,.
\end{align}
Hence, the diffusion of 1D plasma waves proceeds according to
\begin{align}
	\begin{split}
		\mathbf{B}_{\rm 1}\left(x,y,z,t\right)=\:&\mathbf{B}_{\rm 1i}\left(x,y,z,t\right)e^{-\mathfrak{D}t}
	\end{split}\label{eq:ansatzdamping},
\end{align}
where $\mathbf{B}_{\rm 1i}$ denotes the magnetic field of the ideal solution of wave 
dynamics in the absence of any numerical dissipation. The diffusion of the electric field 
$\mathbf{D}_1$ can be modeled analogously.

In order to evaluate the diffusive properties of our code, we define the electromagnetic 
energy contained in the wave components ($\mathbf{B}_1$ and $\mathbf{D}_1$)
\begin{align}
	U_1=\frac{1}{2}\int\text{d}V \left[\mathbf{B}_1^2+\mathbf{D}_1^2\right]\equiv U_{\rm 1i}\,e^{-2\mathfrak{D}t}.\label{eq:freeenergy}
\end{align}
Since the energy $U_1$ is associated with the perturbations of the background electromagnetic 
field, we may refer to it also as free energy.
In our analysis, we evaluate the right-hand side of Eq.~(\ref{eq:freeenergy}), for 
which we obtain the linear relation
\begin{align}
	\ln U_{1}=-2\mathfrak{D}t+\ln U_{\rm 1 i}, \label{eq:logUp}
\end{align}
where $U_{\rm 1 i}$ is the initial energy in the respective waves. Eq.~(\ref{eq:logUp}) 
allows us to obtain $\mathfrak{D}$ (and hence, $\eta_*$) from the slope of linear fits of the 
form $\ln U_1$ versus $t$. The values of $\eta_*$ computed from the fits depend on the grid spacing 
with which the model is run or, equivalently, on the number of points per wavelength, $p=N/m$ ($N$ is the number of points in the domain and $m$ is the number of full modes), 
with which a target mode is resolved in a given setup. Fixing the remaining parts of the numerical 
method, the value of $\eta_*$ also depends on the cell interface reconstruction employed. 
Following \citet[][Sect.~4.3.3]{Rembiasz2017}, we assume that the numerical 
resistivity can be written in the form
\begin{align}
	\eta_*=\mathfrak{N}\times \mathcal{V}\times\mathcal{L}\times \left(\frac{\Delta x}{\mathcal{L}}\right)^{r},\label{eq:NumResistivityDx}
\end{align}
where $\mathfrak{N}$ is a (resolution-independent) numerical coefficient, $r$ the measured order of convergence of the employed scheme, $\Delta x=L/N$ the grid spacing, and $\mathcal{L}$ 
and $\mathcal{V}$ are the characteristic length and speed of the problem. In the case of FFE, 
differently from incompressible MHD, the characteristic velocity is the speed of light, 
thus $\mathcal{V}=1$ in our units. 
The characteristic length is equal to the wavelength of the induced modes \citep{Rembiasz2017}, 
namely, $\mathcal{L}=\lambda=2\pi/k=L/m=p\Delta x$. 
Thus, we can rewrite expression~(\ref{eq:NumResistivityDx}) as
\begin{align}
	\eta_*=\mathfrak{N} \times\lambda\times p^{-r}.\label{eq:NumResistivity2}
\end{align}
In view of the previous relation, we define subsets of tests where the cell interface 
reconstruction is fixed and different number of points per wavelength are considered. For 
every subset of tests (i.e., reconstruction method), 
we fit the function
\begin{align}
	\ln \eta_{*}=-r \ln p+ d\label{eq:etafit},
\end{align}
where the fit parameter $d$ (from which we may compute $\mathfrak{N}$) corresponds to
\begin{align}
	d=\ln\left[\mathfrak{N}\times\lambda \right]=\ln\left[\mathfrak{N}\times L\right]-\ln m\label{eq:etafitparameterd}.
\end{align}
In Fig.~\ref{fig:ResistivityWaves} we combine Eqs.~(\ref{eq:etafit}) and~(\ref{eq:etafitparameterd}) 
in order to visualize the quantity
\begin{align}
	\ln \left[m\times\eta_{*}\right]=-r \ln p+ \ln\left[\mathfrak{N}\mathcal{V}L\right]\label{eq:etafitnormalized}.
\end{align}
\begin{table}
	\centering
	\caption{Assessment of numerical resistivity (numerical fit parameters of the data 
		visualized in Fig.~\ref{fig:ResistivityWaves}, see Eq.~\ref{eq:etafitnormalized}) 
		for a 1D standing sine wave (series A) as well as propagating 1D plasma waves 
		(series B and C, i.e., fast and Alfv\'{e}n). In parentheses we show the results for the cases that differ significantly when using the standard fourth-order reconstruction for the parallel current.}
	\label{tab:resistivity_waves_table}
	{\renewcommand{\arraystretch}{1.5}
		\begin{tabular}{cccc}
			\hline
			Series & Reconstruction & $\mathfrak{N}$ & r \\*[0pt]\hline
			A & MP5 & $34.96$ & $4.88$ \\
			& MP7 & $259.13$ & $6.84$ \\
			& MP9 & $1992.03$ & $8.79$ \\\hline
			B & MP5 & $34.96$ & $4.88$ \\
			& MP7 & $255.34$ & $6.83$ \\
			& MP9 & $783.37$ & $8.38$ \\\hline
			C & MP5 & $36.47$     & $4.87$       \\
			& MP7 & $269.91$     & $6.83$      \\
			&  & (39.94) & (5.8) \\
			& MP9 & $1822.52$     &   $8.73$    \\
			&& (4.29) & (5.17) \\\hline
	\end{tabular}}
\end{table}

We therefore combine measurements of different wave modes $m$ in the same numerical domain 
as a measure of numerical resistivity. For the entire set of considered cases (series A/B/C) 
we find that at least eight grid points per wave mode are required in order to resolve the 
respective plasma wave with an order of convergence which approaches the theoretical order of 
the employed reconstruction scheme. Table~\ref{tab:resistivity_waves_table} shows the 
coefficients $\mathfrak{N}$ and $r$ obtained by a linear fit to Eq.~(\ref{eq:etafitnormalized}). 
For MP5 their values are similar (as expected from Eq.~\ref{eq:NumResistivity2}) 
independent of the series. The numerical diffusivity in series B is somewhat larger compared to series A and C (the last of which uses the eighth-order accurate discretization of $\mathbf{j}_{||}$) for the set of models using the MP9 reconstruction. We have not 
been able to identify a single, dominant source for the slight reduction of the order 
(and the slightly increased diffusivity) of MP9 when applied to fast waves. However, the 
$\lesssim 5\%$ reduction of $r$ does not seem to be a great drawback for applying MP9 in combination with either 4th or eighth-order accurate discretization of $\mathbf{j}_{||}$ to numerical models involving fast waves.

In the right panel of Fig.~\ref{fig:ResistivityWaves}, we also show the effects of the 
order of discretization of the derivatives appearing in the parallel current (Sect.~3.4, Paper I). 
The thick (opaque) lines show the results when a fourth-order accurate discretization is employed 
in combination with MP7 (blue thick line) or MP9 (black thick line) reconstruction. For comparison, 
the blue (black) thin lines show the fits to the resolution dependence of the numerical resistivity 
using a sixth(eighth)-order accurate parallel current discretization in combination with MP7 
(MP9). In order to obtain an empirical order of convergence close to the theoreticaly expected 
values ($r=7$ and 9 for MP7 and MP9, respectively), increasing the order of the discretization 
of $\mathbf{j}_{||}$ is crucial. We note that for resolutions of fewer than eight points per wavelength, 
the values of the numerical resistivity are not too different. This justifies our choice of a 
fourth-order accurate discretization of $\mathbf{j}_{||}$ \citep{Mahlmann2019,Mahlmann2020} 
in combination with MP7. The resolution employed in these global 3D models is smaller than $p\sim 8$ 
zones per wavelength. However, future models with higher resolution will have to incorporate, 
at least, a sixth(eighth)-order accurate discretization of the current when used in 
combination with MP7 (MP9). 

We therefore conclude that modeling force-free plasma wave interactions requires resolving 
the shortest wavelength mode with more than $p=8$ ($p=20$) grid zones in order to reduce
the numerical resistivity below $\sim 10^{-4}$ ($\sim 10^{-7}$). In this way, we avoid significant wave damping over sufficiently long timescales. The typical timescale for energy damping in FFE waves is $\tau\approx 1/(2\mathfrak{D}) \approx 1/(\eta_* k^2)$. In a domain of size $L\sim 1$, where we fit $m\sim 10$ full waves, $\tau \sim 2.5\times 10^{-4}/\eta_*$. Hence, we need a numerical resistivity smaller than $ 2.5\times 10^{-4}$ in order to evolve the set of waves for more than a single domain light crossing time ($t_{\rm lc}=L$) without significant numerical dissipation. That requires $p> 10$.  Applying these 
requirements to 3D numerical experiments (by taking cubical meshes with the aforementioned number of points per wavelength in each dimension) reveals the huge computational effort needed to 
resolve the dynamics of the interaction of FFE modes with sufficiently small numerical
resistivity. This happens because, usually, in global 3D simulations (e.g., of magnetospheres around magnetars), the wavelength of Alfvén waves, $\lambda$, is much smaller than the typical domain size $L$. If, say, $L/\lambda\sim 100$, one may require at least $\sim 1000$ cells per dimension to prevent numerical damping of the waves before they interact for a time on the order of a single light crossing time of the computational domain. If longer evolutionary times are considered, even more stringent numerical resolutions are required.

Figure~\ref{fig:ResistivityOscillations} gives an illustrative example of mode damping for 
different resolutions per wavelength, including the superposition of two selected modes of 
fast waves with the same initial amplitude (adding the initial data in 
Eq.~\ref{eq:fastwavesetup} for $m=1$ and $m=2$). The damping for all cases proceeds 
exponentially according to the assumption we stated in Eq.~\eqref{eq:ansatzdamping} 
with numerical resistivities inferred from Fig.~\ref{fig:ResistivityWaves}. This holds 
also for the case of mode superposition, where the damping of the $m=2$ mode is dominant 
due to its capture by fewer points per wavelength as compared to the $m=1$ mode. The results 
for propagating FFE waves highlight the ability of our code to properly resolve their linear 
superposition with decreasingly small numerical resistivity as the numerical resolution is 
increased.

\subsection{Tearing modes}
\label{sec:TearingModes}

\begin{figure}
	\centering
	\includegraphics[width=0.49\textwidth]{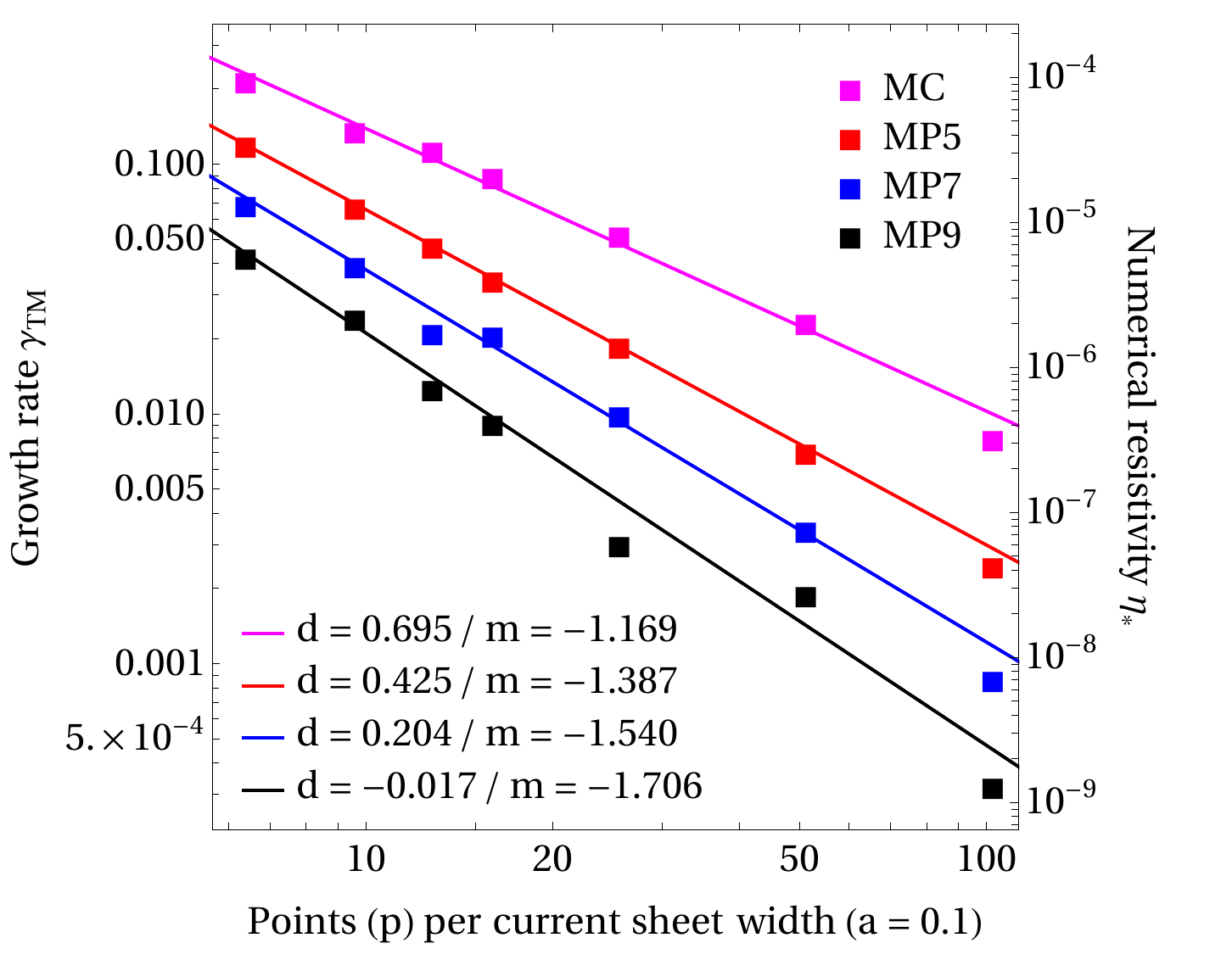}
	\caption{Growth rate of the (2D) TM instability (Sect.~\ref{sec:TearingModes}) 
		for different resolutions and reconstruction schemes. We depict the measurements of 
		$\gamma_{\textsc{tm}}$ from numerical experiments (squares) and the corresponding best 
		fit parameters (solid lines, Eq.~\ref{eq:loggammavslogdx}). The scale on the right 
		shows the numerical resistivity computed from the measurements of the growth rate 
		employing model A (Eq.~\ref{eq:gammatm_2}).}
	\label{fig:TMGrowth}
\end{figure}
\begin{figure}
	\centering
	\includegraphics[width=0.49\textwidth]{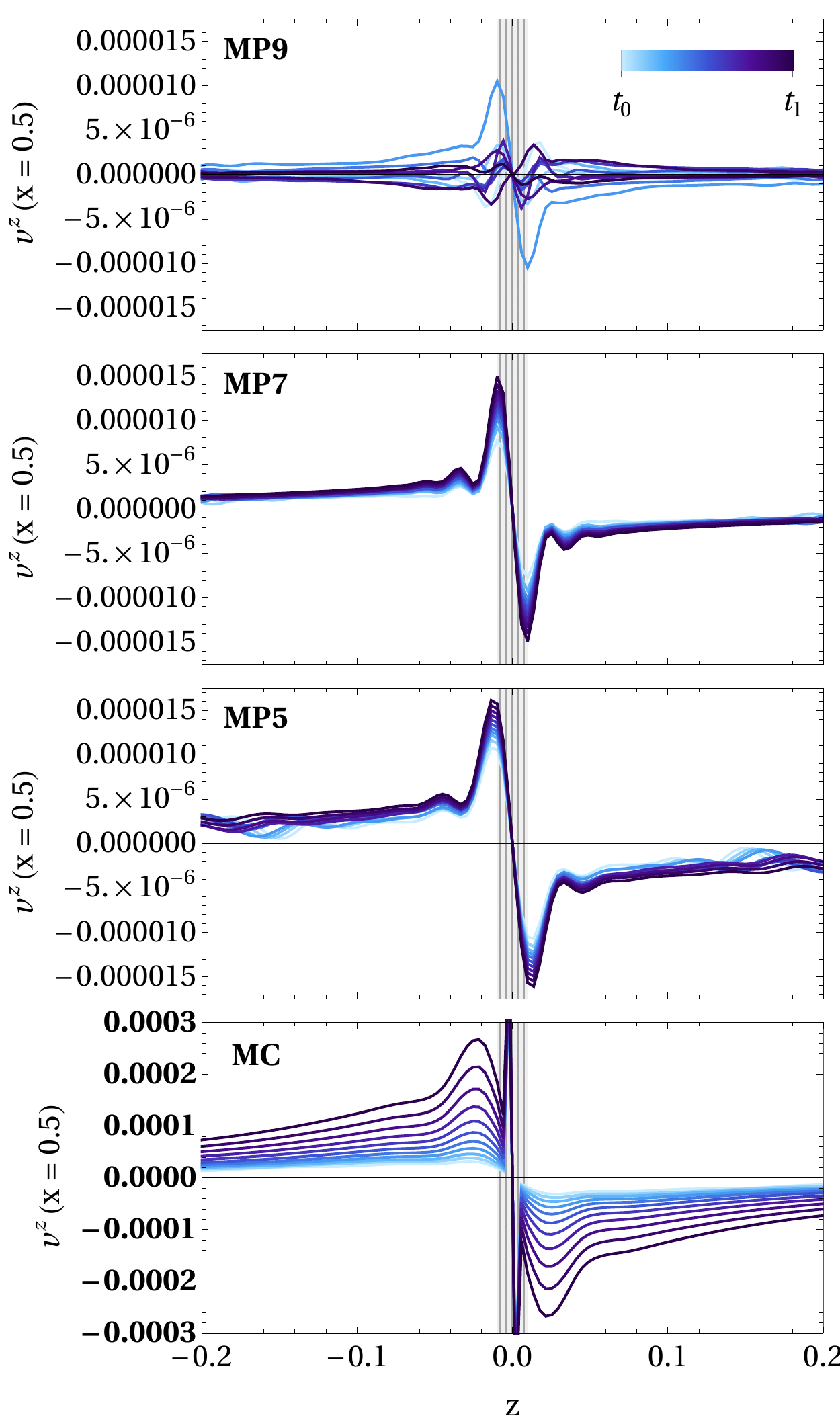}
	\caption{Evolution of the drift velocity $V^z$ along the mid-point of the current 
		sheet ($x=0.5$) in the direction transverse to the current sheet ($z$-direction) 
		during the linear phase (time-interval $\left[t_0,t_1\right]$ for which the growth 
		rate is derived in Fig.~\ref{fig:TMGrowth}). The different panels correspond to different 
		reconstruction schemes (see legends) and a resolution of $p=25.6$ numerical zones per 
		transverse size of the current sheet. The assumed width of the resistive layer 
		$L_{\rm R}$ employed for the analysis in Sect.~\ref{sec:TearingModes} is indicated by 
		the vertical gray lines in the background; each of these lines denotes the limits of the 
		computational zones in the $z$-direction. The growth of the velocity component $V^z$ is 
		strongly suppressed for higher-order (MP) reconstruction methods and is significantly 
		larger when using the MC reconstruction (we note the different scale for the lower panel).}
	\label{fig:VelocityProfile}
\end{figure}
\begin{table}
	\centering
	\caption{Estimates of the dimensionless coefficient $\mathfrak{N}$ and the empirical order 
		of convergence $r$ for the two models defined by Eqs.~(\ref{eq:gammatm_2}) 
		and~(\ref{eq:gammatm_3}). The parameters are obtained by the linear fit given by 
		Eq.~\eqref{eq:loggammavslogdx}. In the last column, we display the relative deviation 
		of the fit parameter $m$ (see Fig.~\ref{fig:TMGrowth}) from its analytical value $m_{\rm a}$ 
		\citep[for the latter, employing a resolution-dependent length scale $\mathcal{L}$, 
		Eq.~\ref{eq:LR}, and ][]{Rembiasz2017} in the corresponding limit and assuming ideal
		order of accuracy of the scheme (see Sect.~\ref{sec:Discussion}).}
	\label{tab:tearing_table}
	{\renewcommand{\arraystretch}{1.5}
		\begin{tabular}{ccccc}
			\hline
			Limit & Reconstruction & $\mathfrak{N}$ & r & $\Delta m/m$ \\*[0pt]\hline
			Model A & MC & $4.4\times 10^{-3}$ & $1.95$ & $0.27$ \\
			& MP5 & $1.2\times 10^{-3}$ & $2.31$ & $0.17$ \\
			& MP7 & $4.7\times 10^{-4}$ & $2.57$ & $0.20$ \\
			& MP9 & $1.7\times 10^{-4}$ & $2.84$ & $0.25$ \\\hline
			Model B & MC & $5.1\times 10^{-3}$ & $2.34$ & $0.38$ \\
			& MP5 & $1.1\times 10^{-3}$ & $2.78$ & $0.28$ \\
			& MP7 & $3.5\times 10^{-3}$ & $3.08$ & $0.3$0 \\
			& MP9 & $1.0\times 10^{-3}$ & $3.41$ & $0.34$ \\\hline
	\end{tabular}}
\end{table}

TM instabilities are resistive instabilities, which can develop in current sheets, and 
dissipate magnetic energy into kinetic energy if a plasma fluid is considered. We note, however, 
that in a pure force-free approximation, the only existing energy is that of the electromagnetic 
field, and, thus, the dissipation of magnetic energy may simply result in an actual energy sink. 
In GRFFE, besides the standard sources of numerical diffusivity, there is yet another one, 
namely the numerical resistivity induced by the algorithms used to control the violations 
of the force-free conditions. As we shall see, both the standard sources of numerical resistivity 
as well as the resistivity produced by the violation of the force-free constraints are especially 
sensitive to the resolution of the numerical mesh and provide a source of (numerical) resistivity from which TMs may develop.

In this test, we adapt the relativistic resistive MHD current sheet setup of \citet{DelZanna2016} 
to FFE, employing a 2D domain of $(x,z) \in \left[-L/2,L/2\right]\times\left[-25a,25a\right]$. 
Here, $a=0.1$ is the width of the current sheet, and $L=2\pi/k$ is its length. In FFE, the 
theoretical growth rate of TMs is analogous to that in incompressible MHD 
\citep[e.g.,][]{Komissarov_etal_2007MNRAS.374..415}. For the wavenumber of the perturbation 
we employ $k=\pi$. This value allows us to fulfill approximately $L\gg a$ and $ka \ll 1$, 
required for the development of TMs. We use periodic boundary conditions in the $x$ direction 
(i.e., along the current sheet) and copy boundary conditions in the $z$ direction (i.e., in 
the direction perpendicular to the current sheet). The numerical grid is uniform and consists 
of $N_x \times N_z$ zones, where $N_x=32$ in all cases. In order to trigger TM instabilities, 
the initial magnetic field
\begin{align}
	\begin{split}
		B_{0x}&= B_0\tanh\left(z/a\right),\\
		B_{0y}&= B_0{\hspace{2pt}\rm sech}\left(z/a\right),
	\end{split}
\end{align}
is perturbed by
\begin{align}
	\begin{split}
		B_{1x}&= \epsilon \left(ak\right)^{-1}B_0\sin\left(kx\right)\tanh\left(z/a\right){\hspace{1pt}\rm sech}\left(z/a\right),\\
		B_{1z}&= \epsilon B_0\cos\left(kx\right){\hspace{1pt}\rm sech}\left(z/a\right).
	\end{split}
\end{align}
We set $B_0=1$ and the perturbation amplitude parameter $\epsilon=10^{-4}$. We assume that 
the initial electric field and the charge density are zero. The growth rate of the TMs, 
$\gamma_\textsc{tm}$, may be traced, for example, by examining the growth of the magnetic field 
component $B_z$ \citep[cf.][]{Rembiasz2017}, which grows exponentially, $B_z = B_z(t=0) e^{\gamma_\textsc{tm} t}$. 
To obtain a globally and positively defined quantity for the growth rate, we study the integral 
of $B_z^2$ over a suitably chosen patch of the computational domain (covering the entire length 
and width of the current sheet): 
\begin{align}
	\ln\int B_z^2 \text{d}S= 2\gamma_{\textsc{tm}}t+\ln \int B_{1z}^2\text{d}S\label{eq:BZGrowth}.
\end{align}
The slope $2\gamma_\textsc{tm}$ of the linear relation in Eq.~(\ref{eq:BZGrowth}) may be obtained 
by fitting $\ln\int B_z^2 \text{d}S$ versus $t$. In the fits we disregard the initial 
(numerically dominated) adjustment phase. We measure the growth rate for series of simulations 
with different numerical resolutions and numerical schemes. Apart from the MP reconstruction 
schemes used in the previous sections we also test slope limited TVD reconstruction 
with a monotonized central (MC) limiter~\citep{vanLeer1977}.

We aim to provide an estimate of the numerical resistivity as a function of the numerical 
resolution based on two different approximations. The first one (model A) assumes that the 
plasma is inviscid. In this case, the growth rate of the TM mode (for a given mode $k$) 
depends on the physical resistivity as \citep[][Eq.~147]{Rembiasz2017}
\begin{align}
	\gamma_{\textsc{tm}}= 1.06^{-4/5}\times \eta^{3/5} v_A^{2/5}a^{-8 / 5}(ak)^{2/5}\left(\frac{1}{a k}-a k\right)^{4 / 5}.
	\label{eq:gammatm_1}
\end{align}
In our system of units, the Alfv\'{e}n speed is $v_{\rm A}=1$. For the specific choice of 
$k$ and $a$ employed in our setup, we can write Eq.~(\ref{eq:gammatm_1}) as 
\begin{align}
	\gamma_{\textsc{tm}}(k=\pi,a=0.1) \approx 55.6\times\eta^{3/5} . 
	\label{eq:gammatm_2}
\end{align}
Alternatively, in the so-called long-wavelength approximation (model B, characterized by 
$ka\ll 1$), the maximum growth rate (i.e., the growth rate of the fastest-growing mode) 
can be evaluated from Fig.~3 of \citet{Furth_etal:1963}, resulting in
\begin{align}
	\gamma_{\textsc{tm}\rm , max}\approx0.6\times a^{-3/2}v_A^{1/2}\eta^{1/2}
	\label{eq:gammatm_0}
\end{align}
or, equivalently,
\begin{align}
	\gamma_{\textsc{tm}\rm , max}(a=0.1) \approx 19.0\times \eta^{1/2} . 
	\label{eq:gammatm_3}
\end{align}
The force-free models we run do not include any physical resistivity. Hence, the growth of 
TM modes is induced by the action of the resolution-dependent numerical resistivity $\eta_*$. 
Thus, we may replace $\eta$ in Eqs.~\eqref{eq:gammatm_2} and \eqref{eq:gammatm_3} by $\eta_*$. 

Once the TM growth rate is obtained for each resolution and reconstruction method, we can 
compute the corresponding numerical resistivity 
\citep[as we did in Eq.~\ref{eq:NumResistivityDx}, cf.][Sect.~4.3.3]{Rembiasz2017}. 
Like in the previous section, the light speed ($\mathcal{V}=1$) is the only possible choice 
for the characteristic velocity of the problem in the force-free regime. The selection of 
$\mathcal{L}$ is much more involved \citep[see][for a detailed discussion]{Rembiasz2017} and 
prone to accuracy restrictions since typically $\mathcal{L}\ll a$, which can result in a prohibitive numerical resolution needed to reliably measure the growth rate. To obtain an order 
of magnitude estimate, we may assume that
\begin{align}
	\mathcal{L}\sim 0.1a,
	\label{eq:Lcharacteristic}
\end{align}
but cautiously note that assuming that $\mathcal{L}$ is constant, in other words resolution-independent, 
is not fully accurate \citep{Rembiasz2017}. A partial justification for this choice 
(but see discussion in Sect.~\ref{sec:Discussion}) is the shape of the drift velocity, 
$\mathbf{V} = \mathbf{D}\times\mathbf{B}/B^2$, displayed in  Fig.~\ref{fig:VelocityProfile},
which shows that $0.1 a$ (region with gray lines) covers a significant part of the width 
of the current sheet (estimated as the distance between the two extrema).

We express the numerical resistivity in terms of the number of grid zones within the transition 
layer, namely, $p=a/\Delta z$ ($\Delta z$ being the grid spacing in the $z$ direction). We note 
that $p$ here has a different meaning as in Sect.~\ref{sec:PlasmaWaves}, and the numerical 
resistivity (Eq.~\ref{eq:NumResistivityDx}) in these tests can be written as (compare with 
Eq.~\ref{eq:NumResistivity2})
\begin{align}
	\eta_*=\mathfrak{N}\times 0.1 a\times \left(0.1p\right)^{-r},\label{eq:NumResistivity}
\end{align}
where $\mathfrak{N}$ is, again, a (resolution-independent) numerical coefficient. 
Plugging Eq.~\eqref{eq:NumResistivity} into either Eq.~\eqref{eq:gammatm_2} or Eq.~\eqref{eq:gammatm_3} 
one obtains a linear relation between $\ln\gamma_\textsc{tm}$ and $\ln p$ that allows us to 
compute $\mathfrak{N}$ and $r$ from the coefficients of the linear fit 
\begin{align}
	\ln\gamma_\textsc{tm}=m\ln p+d,
	\label{eq:loggammavslogdx}
\end{align} 
where the specific definition of $m$ and $d$ depends on the employed growth model. 

Figure~\ref{fig:TMGrowth} visualizes the growth rates $\gamma_{\textsc{tm}}$ for different 
resolutions and reconstruction methods. We summarize the numerical fit parameters, 
namely, the coefficient $\mathfrak{N}$ and the estimated order of the scheme $r$ in 
Table~\ref{tab:tearing_table}. The order of convergence estimated from the fits is roughly 
correct for models using the MC reconstruction, but is significantly smaller than the 
formal order of convergence estimated for MP methods, where values of $r=5$, $7$ and $9$ 
are expected.
In fact, we expect that the methodology employed in this test may only allow us to infer 
order of magnitude estimates of the numerical resistivity and rough values of the empirical 
order of convergence of the algorithm, $r$. 

The reason for the discrepancy is related to the nature of the TM itself, which 
involves dissipation in a current sheet. Dissipative effects do not exist in FFE (except numerically)
and the presence of these current sheets breaks the force-free condition itself (see the 
discussion in Sect.~\ref{sec:Discussion}). In this regime, it is not surprising that our 
numerical methods do not behave as expected and, hence, 
the theoretical order of convergence is not recovered.
These circumstances are not present in the tests in Sect.~\ref{sec:PlasmaWaves} where no 
current sheets are on hand and no reduction of the expected order of convergence is observed.  
A possible solution to this problem is to include a physically consistent model for the 
dissipation in FFE, such that TMs can develop in a physically realistic way forming resistive 
layers resolvable by numerical simulations. Its application to different astrophysical scenarios 
(where the resistivity is likely too small to be handled) would be done by studying the  
limit of decreasing resistivity. The next section explores this possibility.

The influence of the exact boundary conditions on the growth rate in the direction transverse 
to the current layer is probably significant and deserves further study, something that is 
beyond the scope of this publication. We have considered different values of $a$, finding 
that $a=0.1$ is a compromise to fulfill all the physical requirements to develop TMs and 
the (unfortunately unsuccessful) attempt to resolve the resistive layer width. However, 
we stress the fact that our results are numerically robust in several ways. For instance, 
we have checked the influence of the computational box size in the direction perpendicular 
to the current sheet. We find that boxes extending in the $z$-direction by less than $\pm 20a$ 
induce significant modifications in the growth rate of TMs. Our choice of a box size 
extending up to $z=\pm 25a$ was validated by cross-checking against even larger 
computational boxes and finding no appreciable changes in the growth rate. We have also 
considered higher-order discretizations of the derivatives used to compute $\mathbf{j}_{||}$ 
(Sect.~3.4, Paper I), finding no significant influence in this test, likely because of the 
non-smoothness of the parallel current at $z=0$. The discontinuity of the field derivatives 
makes a high-order discretization of $\mathbf{j}_{||}$ less effective (i.e., noisier). 
This result is consistent with our finding that higher-order discretization of $\mathbf{j}_{||}$ 
does not significantly change the results of the 1D Riemann problems presented in Sect.~4.1 
of Paper I.

\section{Beyond ideal force-free electrodynamics}
\label{sec:Beyond_Ideal}

FFE is inherently formulated in the limit of infinite conductivity, effectively resulting 
in the magnetic dominance and ideal fields condition \citep[see Paper I and also][]{Lyutikov2003}
\begin{align}
	\mathbf{B}^2 -\mathbf{D}^2&\ge 0,\\
	\mathbf{D}\cdot\mathbf{B}&=0.
\end{align}
Under these conditions the 4-vector electric current density takes the form of the force-free 
current (Eq.~47 in Paper I):
\begin{align}
	\begin{split}
		I^\mu = I^\mu_\textsc{ff}&=\rho n^\mu+\frac{\rho}{\mathbf{B}^2}\eta^{\nu\mu\alpha\beta}n_\nu D_\alpha B_\beta\\
		&+\frac{B^\mu}{\mathbf{B}^2} \:\eta^{\sigma\alpha\beta\lambda}n_\sigma\left(B_{\lambda; \, \beta}B_\alpha-D_{\lambda; \, \beta}D_\alpha\right),
		\label{eq:FFCurrent}
	\end{split}
\end{align}
where $\rho = -n_\mu I^{\mu} = \alpha I^t$ is the charge density, 
$n^{\mu} =\alpha^{-1}(1, \beta^i)$ is the vector normal to the hypersurfaces in the 3+1 
decomposition of the spacetime, $\alpha$ is the lapse function and $\beta^i$ is the shift 
vector, respectively. 
Equation~(\ref{eq:FFCurrent}) is a direct consequence of 
\begin{equation}
	\mathcal{L}_{n}  (\mathbf{B}\cdot\mathbf{D})=n^{\lambda}\nabla_{\lambda} (\mathbf{B}\cdot\mathbf{D}) =0,
	\label{eq:CurrentDriverFFE}
\end{equation}
where $\mathcal{L}_{n} $ is the Lie derivative with respect to $n^\mu$. The current density 
three-vector appearing in the 3+1 decomposition of Maxwell's equations can be computed as  
$J^i = \alpha I^i$. Additionally, one can define the current density 3-vector for the normal 
observer ($n^\mu$) as the projection of $I^{\mu}$ onto the hypersurface, namely 
$j^{\, i} = I^i - \beta^i I^t$, which is related to $J^i$ by $J^i =\alpha j^{\, i} - \beta^i \rho$.

Any occurring resistivity measured in such ideal schemes must, hence, be of numerical origin; 
we quantify this numerical resistivity for our method in Sect.~\ref{sec:Numerical_Resistivity}. 
The absence of any physical resistivity makes resolving genuinely resistive layers a true limit of 
FFE. Thus, it is desirable to extend the theory to include a small phenomenological resistivity 
$\eta$ and reduce to FFE in the limit $\eta\rightarrow0$. Several attempts to undertake this 
task have been presented in the literature \citep[e.g.,][]{Lyutikov2003,Gruzinov2007,Li2012,Parfrey2017}. 
In spite of the oxymoron, it is common to refer to 
these prescriptions as resistive FFE. All of them have in common that they integrate the 
full set of Maxwell's equations together with a suitable choice of Ohm's law. In practice, 
this replaces the force-free electric current density $I^\mu_\textsc{ff}$ by a more general 
form explicitly including the resistivity. For this section, we have tested the generalization 
of a current density recently introduced by \citet{Parfrey2017}. Its covariant form 
(cf. Eq.~\ref{eq:FFCurrent}) is
\begin{align}
	\begin{split}
		I^\mu &=\rho n^\mu+\frac{\rho}{\mathbf{B}^2}\eta^{\nu\mu\alpha\beta}n_\nu D_\alpha B_\beta\\
		&+\frac{1}{1+\kappa_I\eta}\frac{B^\mu}{\mathbf{B}^2} \left[\:\eta^{\sigma\alpha\beta\lambda}n_\sigma\left(B_{\lambda; \,\beta}B_\alpha-D_{\lambda; \,\beta}D_\alpha\right)+\kappa_I B^\sigma D_\sigma\right],
		\label{eq:FFResCurrent}
	\end{split}
\end{align}
where $\kappa_I$ is a constant and $\eta$ is the resistivity. With this new 4-current the 
evolution of $\mathbf{B}\cdot\mathbf{D}$ fulfills \citep[cf.][]{Parfrey2017}
\begin{align}
	\mathcal{L}_{n}  (\mathbf{B}\cdot\mathbf{D})& =\kappa_I\left(\alpha^{-1}\eta\, \mathbf{J}-   \mathbf{D} \right)\cdot \mathbf{B}\nonumber \\
	&=\kappa_I\left(\eta\, \mathbf{j}_{||}-   \mathbf{D}_{||} \right)\cdot \mathbf{B} - \kappa_I  \eta \alpha^{-1} \rho\: \pmb{\beta}\cdot\mathbf{B},
	\label{eq:CurrentDriver}
\end{align}
where  $||$ indicates the component of a vector parallel  to $\mathbf{B}$.
Therefore, $1/\kappa_I$ is the timescale over which the parallel electric field $\mathbf{D}_{||}$  
is driven toward $\alpha^{-1}\eta\,\mathbf{J}_{||}$ (or in for the case of flat spacetime, 
toward $\eta\,\mathbf{j}_{||}$). Therefore, the form of this current effectively is enforcing 
a form of Ohm's law for the parallel current. In the limit $\eta\to 0$ one recovers 
$\mathbf{B}\cdot\mathbf{D} \to 0$ and hence $I^\mu \to I^\mu_\textsc{ff}$.

The current density observed by the normal observer naturally splits into components 
parallel and perpendicular to the magnetic field three-vector 
\citep[$\mathbf{j}=\mathbf{j}_\perp + \mathbf{j}_{||}$, cf. Eq.~70 in][]{Komissarov2011}:
\begin{align}
	\mathbf{j}_{\perp} &= \rho \frac{\mathbf{D} \times \mathbf{B}}{\mathbf{B}^2},\\
	\mathbf{j}_{||} &= \frac{1}{1 + \kappa_I\eta} \frac{  \mathbf{B} \cdot(\nabla\times\mathbf{B})  - 
		\mathbf{D}\cdot(\nabla\times\mathbf{D}) + \kappa_I \: \mathbf{B}\cdot \mathbf{D}}{\mathbf{B}^2} \:\mathbf{B} 
	\label{eq:FFResCurrentPerpendicular}.
\end{align}

In the following subsections, we will investigate the ability of the current given by 
Eq.~(\ref{eq:FFResCurrent}) to model resistivity in FFE. We specifically address its 
ability to a) resolve current carrying discontinuities in 1D plasma wave tests 
(extending Sect.~4.1 of Paper I) as well as b) its capacity to resolve the resistive 
layer in 2D TMs (cf. previous Sect.~\ref{sec:TearingModes}). Along the way, 
we cross-validate our procedure to evaluate the numerical resistivity of our algorithm by 
comparing this prescription for resistive FFE with its ideal limit.

From the numerical point of view, the introduction of resistive effects needs to be done 
with extra care. The choice of the driving rate $\kappa_I$ is guided by numerical convenience. 
Its value shall be large enough to drive $\mathbf{D}_{||}$ toward $\alpha^{-1}\eta\,\mathbf{J}_{||}$  
as quickly as possible., however, it shall not be so large as making the driving timescale 
become much smaller than the time step of the numerical model ($1/\kappa_I\ll \Delta t$). 
In this case, Maxwell's equations become numerically stiff and suitable time evolution algorithms 
such as implicit Runge-Kutta methods \citep{Palenzuela2009,Miranda-Aranguren2014IAUS..302...64A,Miranda-Aranguren2018} would need to be employed. Numerical experimentation 
leads us to resort to $\kappa_I \Delta t \sim 1 - 10$. We note that if the modified current is 
employed, the perpendicularity condition $\mathbf{D}\cdot\mathbf{B}=0$ must not be enforced 
anymore. We note that Eq.~(\ref{eq:CurrentDriver}) is actually similar to the evolution equation 
used by \cite{Alic2012} to drive $\mathbf{B}\cdot\mathbf{D}$ toward zero in 
FFE simulations (with $\eta=0$), an alternative to our approach described in Paper I.

\subsection{Charge-carrying 1D discontinuities}
\label{sec:1D_Charge_Carrying}

\begin{figure*}
	\centering
	\includegraphics[width=0.98\textwidth]{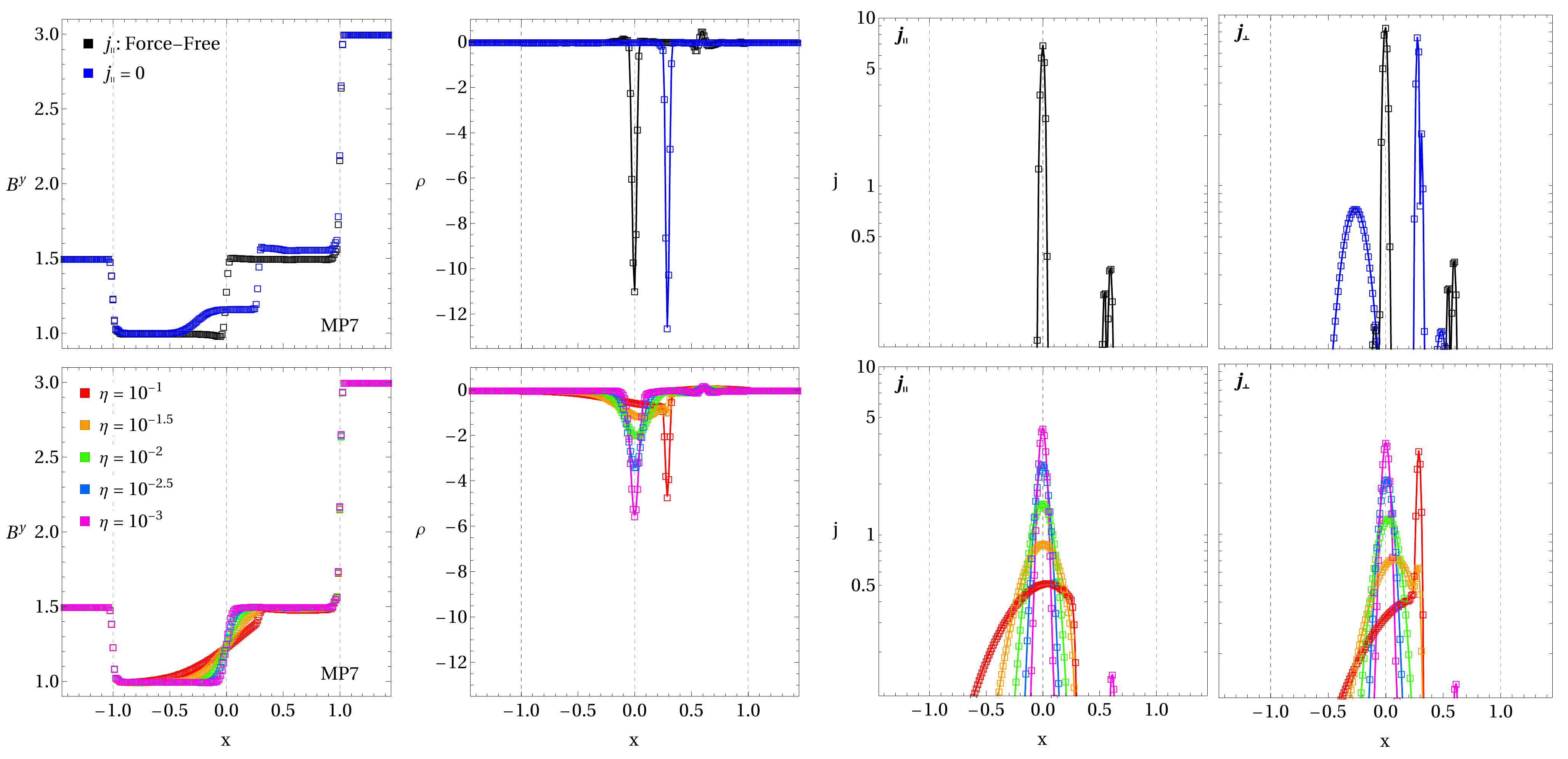}
	\caption{Three-wave test ($\Delta x=0.0125$) employing different Ohm's laws. We display 
		the magnetic field component $B^y$, the electric charge $\rho$, as well as the magnitudes 
		of the parallel ($\mathbf{j}_\parallel$) and perpendicular ($\mathbf{j}_\perp$) projections 
		of the current. \textit{Top panels:} Conventional force-free current $I_{\rm FF}$ 
		(black color, Eq.~\ref{eq:FFCurrent}) and $\mathbf{j}_\parallel=0$ (blue color) 
		with an algebraic enforcement of the force-free conditions. \textit{Bottom panels:}  
		FFE current (Eq.~\ref{eq:FFResCurrent}) for different values of $\eta$.}
	\label{fig:JparJperp}
\end{figure*}

The three-wave test in Sect.~4.1.2 of Paper I is a well-suited testbed for the analysis 
of resistivity introduced by Ohm's law. This test consists of an initial discontinuity that 
splits into two fast discontinuities and one stationary central standing Alfv\'{e}n wave.
The Alfv\'{e}n wave is a charge carrying discontinuity, which is effectively stabilized by 
strong currents. In this section, we compare the results of the three-wave test using three 
different Ohm's laws. In all tests we use the highest resolution employed in Paper I, 
namely, $\Delta x=0.0125$. First, we consider the force-free current given by Eq.~(\ref{eq:FFCurrent}) 
and enforce the force-free conditions algebraically (corresponding theoretically to $\eta=0$), 
as described in Paper I. Second, we set $\mathbf{j}_\parallel=0$ \citep{Yu2011}, again 
enforcing the force-free conditions algebraically. Third, we employ the current density 
given by Eq.~(\ref{eq:FFResCurrent}) (with $\kappa_I=512$) and refrain from algebraically enforcing the perpendicularity condition. In other words, the magnetic dominance condition is always validated algebraically; the different Ohm's laws deviate in their action on $D_\parallel$. 

Figure~\ref{fig:JparJperp} combines test results with the following observations. The force-free current $I_\textsc{ff}^\mu$ preserves the central charge carrying 
	discontinuity well. Setting $\mathbf{j}_\parallel=0$ triggers the splitting of the central 
	discontinuity into two waves: One charge-carrying discontinuity moving to the right, 
	and a shallow layer to the left of the mid-point ($x=0$). The resistive  Ohm's law implementing $I^\mu$ approaches the force-free solution 
	for decreasing values of $\eta$. For $\eta=10^{-1}$, the largest considered value, 
	the charge-carrying discontinuity shifts to the right of the central point. Decreasing 
	the resistivity to smaller values $\eta=10^{-3}$ almost reproduces the ideal FFE results: 
	Both charge and stabilizing currents around the central point gradually increase.

The fact that decreasing values of the phenomenological resistivity $\eta$ approach the results 
from ideal FFE is showing (reassuringly) that the numerical resistivity in our ideal formulation 
of FFE is below $10^{-3}$ even in charge-carrying discontinuities. 
This physical resistivity corresponds to a magnetic Reynolds number of $\mathcal{R}_m=\mathcal{V}\mathcal{L}/\eta = 12.5$, when one considers $\mathcal{V}=1$ and $\mathcal{L}=\Delta x$ (the natural lengthscale $\mathcal{L}$ for this test is the width of the discontinuity, which is numerically of order $\Delta x$). We note that the typical length scale is purely numerical, while the resistivity is physical. Since the Riemann problem we are dealing with is self-similar (at least in the ideal limit $\eta\rightarrow 0$), strictly speaking, there is no physical characteristic length scale. Thus, interpretation of the estimated value of the Reynolds number has to be taken with caution. It simply compares the light-crossing time of a numerical cell, $\Delta t = \Delta x$, with the resistivity (which has dimensions of time in our units), and states that the resistivity is $\sim 10$ times smaller than $\Delta t$ in this test.
The results presented in 
Fig.~\ref{fig:JparJperp} also include somewhat counter-intuitive observations: Setting 
$\mathbf{j}_\parallel=0$ and enforcing the force-free conditions algebraically 
\citep[hence, $\eta\rightarrow 0$,][]{Yu2011} shows a similar behavior as the case of 
$\eta=10^{-1}$ (hence, $\eta\gg 0$) in the resistive modeling. We attribute this contradictory 
behavior to the feature of enhanced charge conservation, which we ensure in our code 
(see Paper I). Abandoning part of the current ($\mathbf{j}_\parallel$) and only enforcing 
$\mathbf{D}\cdot\mathbf{B}=0$ algebraically leads to an inconsistency between the charge 
$\rho$ and the electric field $\mathbf{D}$. As we evolve the continuity equation of the 
charge density (and do not reconstruct $\rho=\text{div}\mathbf{D}$ in every timestep, as 
most other FFE codes do) the algebraic reset of the electric fields to enforce  
$\mathbf{D}\cdot\mathbf{B}=0$ without a consistent modeling of the respective Ohm's law in 
the source terms is no longer adequate. The possible mismatches between charge densities and 
electric fields become obvious in this test and justifies the additional evolution equation 
we include into our scheme to ensure the long-term validity of charge-densities and electromagnetic 
fields in global astrophysical simulations 
\citep[as we found especially useful in][]{Mahlmann2019,Mahlmann2020}.

\begin{figure}
	\centering
	\includegraphics[width=0.49\textwidth]{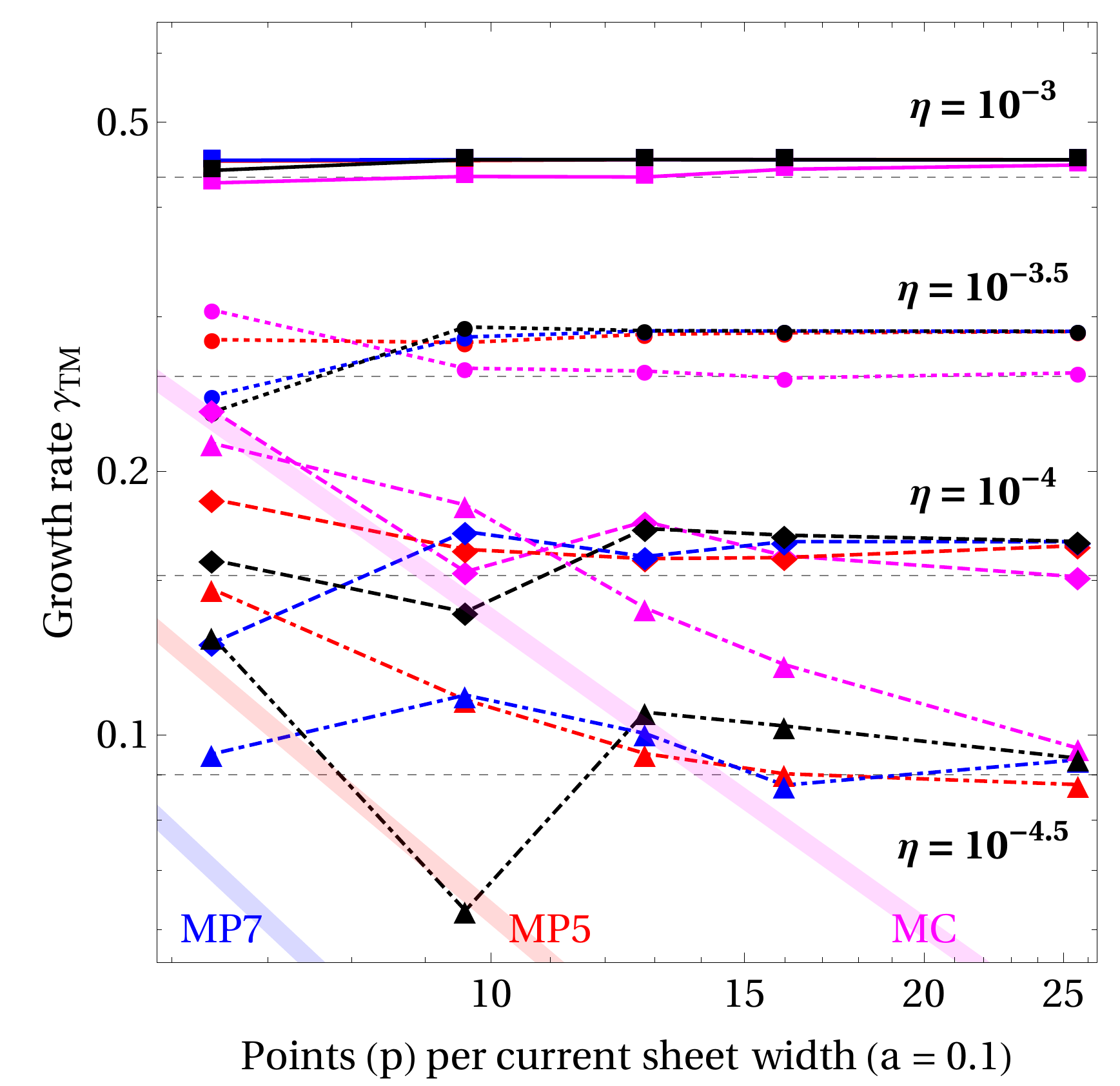}
	\caption{Measured growth rates for simulations using different resistivity parameters 
		$\eta$ in the resistive current density, Eq.~(\ref{eq:FFResCurrent}), and for 
		different reconstruction schemes (MC: magenta, MP5: red, MP7: blue, MP9: black) as a 
		function of the numerical resolution. Thick lines in the background show the fits 
		obtained in Sect.~\ref{sec:TearingModes} for numerical simulations without any 
		added resistivity (see Fig.~\ref{fig:TMGrowth}). As $\eta$ is lowered, the effect 
		of numerical resistivity emerges (especially in the bottom left corner of the panel). 
		Dashed gray lines mark the values of the growth rates corresponding to the values of 
		$\eta$ used in the different series of simulations, computed using 
		Eq.~(\ref{eq:gammatm_2}).} 
	\label{fig:TMResGrowth}
\end{figure}
\begin{figure}
	\centering
	\includegraphics[width=0.49\textwidth]{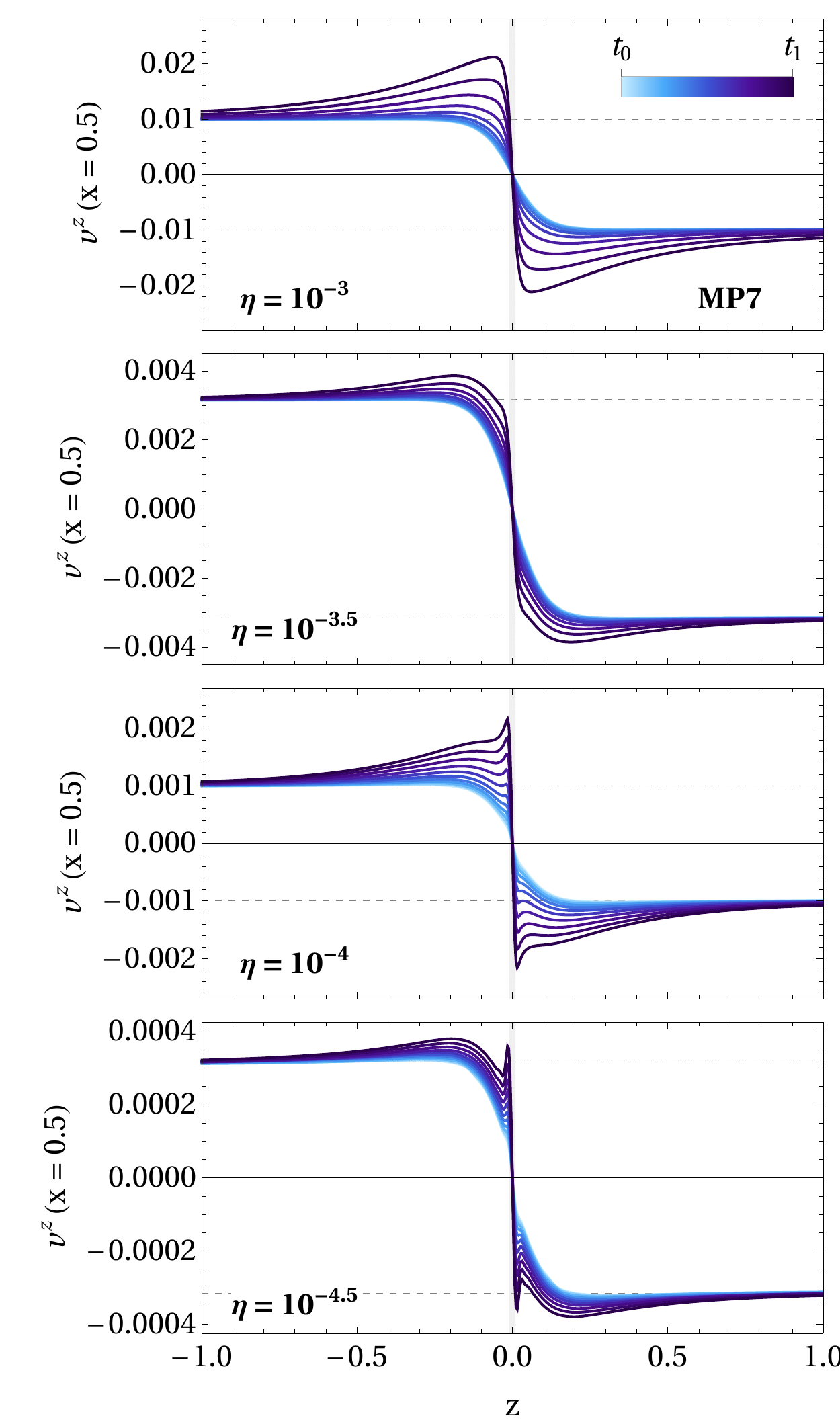}
	\caption{As Fig.~\ref{fig:VelocityProfile} but showing one specific reconstruction 
		method (MP7) for different resistivity parameters $\eta$. The asymptotic velocity is 
		driven to a magnitude $\eta/a$ (dashed gray lines), different from zero, by the isotropic 
		phenomenological resistivity employed in $I^\mu$.}
	\label{fig:VelocityProfileRes}
\end{figure}
%

\subsection{Tearing modes driven by physical resistivity}
\label{sec:TM-physical}

In this section, we repeat the TM tests from Sect.~\ref{sec:TearingModes} for 
the resistive current $I^\mu$ from Eq.~(\ref{eq:FFResCurrent}). In all cases we 
use $\kappa_I=512$. We aim at probing if such phenomenological currents can be used to model 
resistive layers in FFE more accurately, in the limit of decreasing resistivity. The principal 
idea behind this test is that in resistive FFE simulations we have two sources of resistivity. 
Namely, the physical resistivity $\eta$, set in the expression for the current, and the numerical 
resistivity $\eta_*$, depending on the numerical scheme and resolution (as seen in Sect.~\ref{sec:TearingModes}). 
For sufficiently high resolution, one should obtain convergent results depending only on $\eta$. If 
the resolution is decreased, the point at which the growth rate $\gamma_{\textsc{tm}}$ starts 
differing from the converged result should correspond to $\eta_* \sim \eta$, and marks the 
limit at which effects of numerical resistivity become dominant. This allows for an independent 
estimation of the numerical resistivity of the code, to be compared with those in 
Sect.~\ref{sec:TearingModes}.

Figure~\ref{fig:TMResGrowth} assembles growth rates for different values of the resistivity 
parameter $\eta$. If the imposed physical resistivity is well above the numerical resistivity 
quantified in Sect.~\ref{sec:TearingModes}, namely, $\eta_{*}<\eta$, the measured growth rates 
$\gamma_{\textsc{tm}}$ converge to a single value almost independently of the employed resolution 
and reconstruction method. As $\eta$ is reduced approaching the limit set by the numerical 
resistivity, the growth rates become more scattered around the corresponding converged value. 
In cases where numerical and imposed resistivity are comparable, $\eta_{*}\approx\eta$, 
(e.g., for MC reconstruction with the lowest $\eta$) the growth rate can be explained by 
the combined effect of numerical and physical resistivity. Finally, we expect that in case 
of $\eta_{*}>\eta$, the evolution is fully dominated by numerical resistivity.

The solution in the regime with finite resistivity differs in qualitative ways with respect 
to the ideal FFE case. According to \citep{Low_1973ApJ...181..209} the velocity component 
$V^z$ is driven to a finite value $\eta/a$ asymptotically by the employed isotropic resistivity 
model. We observe this feature in Fig.~\ref{fig:VelocityProfileRes}, where the drift velocity 
approaches asymptotically the dashed gray line representing $\eta/a$. This behavior differs 
from the ideal case studied in Sect.~\ref{sec:TearingModes}, where $\mathbf{V}=0$ asymptotically 
(for sufficiently high numerical resolution, as seen in Fig.~\ref{fig:VelocityProfile}). 
The key difference between both cases is that the numerical resistivity measured in 
Sect.~\ref{sec:TearingModes} acts mainly in the region of the current sheet, while in the 
resistive simulations in this section $\eta$ is the same over the whole domain. Therefore, 
the (physical) boundary conditions of both sets of simulations differ slightly 
(although in the limit $\eta \to 0$ they do converge). 

The widths of the resistive layers also differ from the FFE results. In 
Sect.~\ref{sec:TearingModes} (see Fig.~\ref{fig:VelocityProfile}) we found that the 
size of the resistive layer was effectively of the size of a few numerical cells. However, 
in the resistive simulations (see Fig.~\ref{fig:VelocityProfileRes}) the widths of the resistive
layers (estimated as the extent between the extrema of the drift velocity profile) extends 
well beyond a few grid zones off the central current sheet, especially for $\eta>\eta_{*}$. 
In this case, the layer is well resolved. The difference between the two classes of 
simulations is that in resistive FFE the width of the resistive layer is set by the value of 
$\eta$ while in the ideal FFE case its width is set by the numerical resistivity and shrinks 
with increasing resolution. In other words, the velocity should vanish as the numerical resistivity decreases and indeed this is what is seen in Fig.~\ref{fig:VelocityProfile}: for higher reconstruction order there is  smaller resistivity and, hence, smaller drift velocity. Similar behavior is observed for the case with physical resistivity (Fig.~\ref{fig:VelocityProfileRes}) in which, as long as the physical resistivity dominates over the numerical one, the drift velocity decreases with decreasing resistivity. We note however, that close to the current sheet, numerical resistivity starts to dominate toward the lower panels. In conclusion, the width of the shear layer in FFE is a moving target 
as we increase resolution and, depending on the numerical scheme used, it may not be possible to
resolve it with increasing resolution. For cases with $\eta\approx\eta_{*}$, additional 
(local) maxima emerge corresponding to a thinner resistive layer associated with numerical 
resistivity (see bottom panels of Fig.~\ref{fig:VelocityProfileRes}).

The fact that the region of convergent results of $\eta>\eta_{*}$ roughly coincides with the 
limits established in our previous tests further supports our findings from Sect.~\ref{sec:TearingModes}: 
For intermediate resolutions of $p\approx 10$, our GRFFE tool has a numerical resistivity below 
$\eta_{*}\approx 10^{-4}$, corresponding to a magnetic Reynolds number of $\mathcal{R}_m \approx 100$ (estimated by taking the characteristic length scale given by Eq.~\ref{eq:Lcharacteristic}). We stress that other physical length scales could be used to define a magnetic Reynolds number, for example, $L$ or $a$, for which one would find $\mathcal{R}_{m,L}\approx 20000$ and $\mathcal{R}_{m,L}\approx2\times 1000$, respectively. Hence, a magnetic Reynolds number classifies the diffusive regime in which our results are obtained ambiguously at best. 

Phenomenological Ohm's laws as the one introduced in 
Eq.~(\ref{eq:FFResCurrent}) are apt to model resistive effects in FFE. 
Although we cannot expect, for the time being, to be able to work with the values of 
resistivity expected in most astrophysical scenarios, at least we can aim at performing 
simulations with controlled values of the resistivity and study its behavior in the limit of 
vanishing $\eta$. Notwithstanding these encouraging results, using resistive FFE brings its 
own problems, such as the appearance of a non zero asymptotic drift velocity moving away 
from the resistive layer in TMs. This indicates the need for further fine-tuning when 
employing resistive Ohm's laws in FFE. Non-isotropic resistivities 
\citep[as employed, e.g., in][]{Komissarov2004} or current-sheet-capturing models 
\citep[as used in][]{Parfrey2017} are candidates for driving the asymptotic dynamics of current 
sheet instabilities further toward the physical solution. Such fine tuning is very likely 
to depend on the employed method and the problem at hand, namely, the correct determination 
of the threshold $\eta\approx\eta_{*}$. A full implementation of different resistive FFE 
models with nonconstant or non-isotropic resistivities will be further explored in the future. 
Correctly modeling phenomenological Ohm's laws could be a valuable asset when bridging codes 
of different plasma regimes as they allow for physical transition layers 
(and consistent signal propagation).

\section{Discussion}
\label{sec:Discussion}

\begin{figure}
	\centering
	\includegraphics[width=0.49\textwidth]{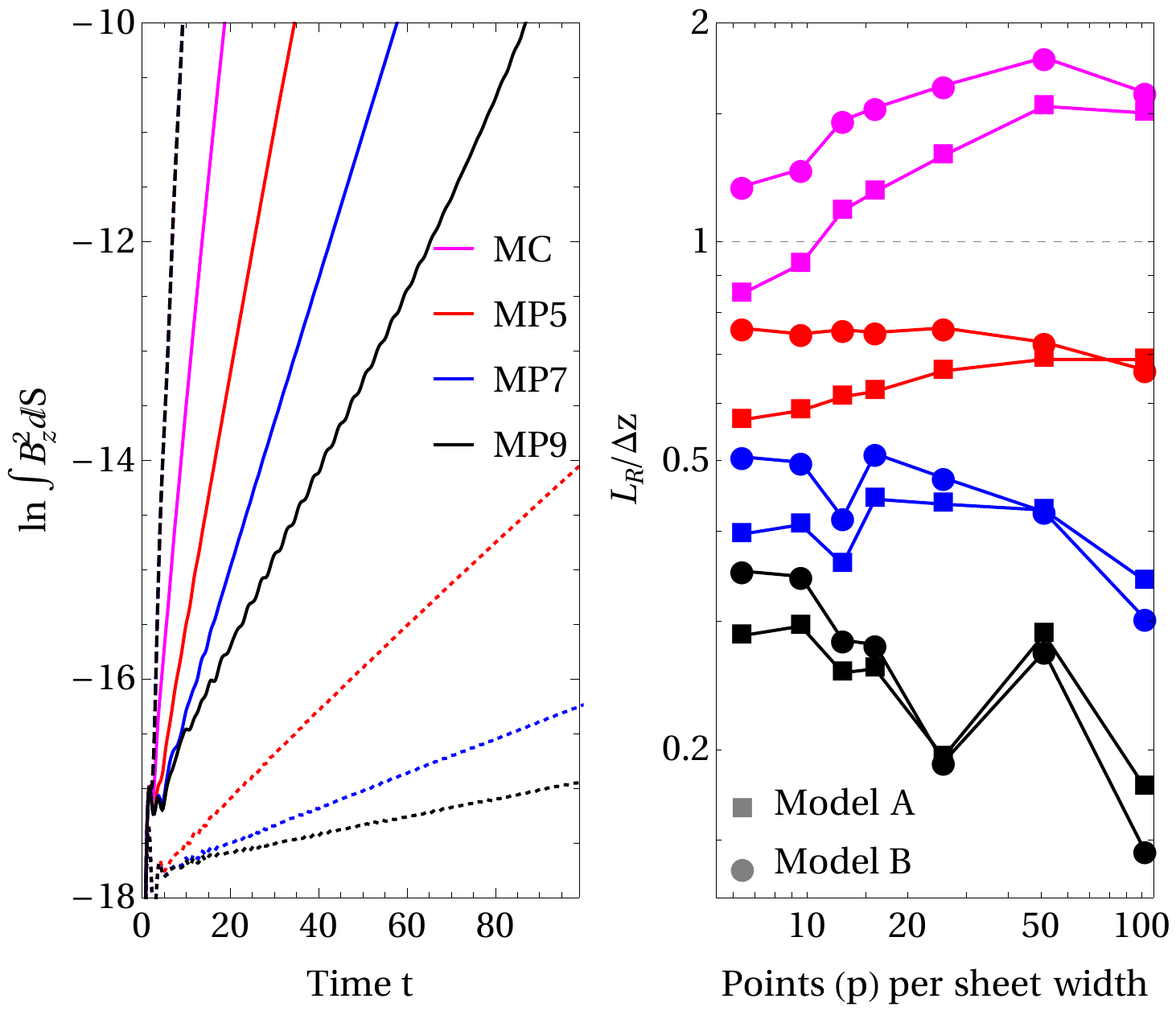}
	\caption{TM growth rates and resistive layer size for different numerical schemes. \textit{Left:} Growth of the (2D) TM instability for different treatments of the parallel current and a fixed number of points per current sheet 
		width $p=6$. We display our standard method for treating $\mathbf{J}_{||}$ (i.e., with a 
		fourth-order accurate discretization of the derivatives, solid lines), extreme cases in 
		which $\mathbf{J}_{||}=\mathbf{0}$ \citep[dotted lines, following][]{Yu2011}, and the 
		case in which the algebraic condition $\mathbf{D}\cdot\mathbf{B}=0$ is not enforced 
		(dashed lines). Different colors correspond to distinct reconstructions (see legends). 
		In the case of no enforcement of the $\mathbf{D}\cdot\mathbf{B}=0$, the results are 
		indistinguishable for the three MP reconstructions. \textit{Right:} Size of the 
		resistive layer $L_{\rm R}$ normalized to the grid spacing $\Delta z$ in the direction 
		perpendicular to the current sheet for all the reconstruction methods (see legends) and 
		models (A, circles and B, squares).}
	\label{fig:TMGrowth2}
\end{figure}

The discretization of any system of partial differential equations (PDEs) on a finite mesh 
of cells to make them amenable for their numerical integration unavoidably introduces 
numerical diffusion and dispersion. The equations of GRFFE are no exception to this rule. 
Furthermore, the necessity of enforcing the conditions specific to GRFFE 
($\mathbf{D}\cdot\mathbf{B}=0$ and $\mathbf{B}^2-\mathbf{D}^2>0$) is an additional source 
of numerical diffusion. The use of monotonic and consistent numerical methods guarantees 
the convergence to the physical non-dissipative solution of a problem in the limit 
of vanishing cell size. In this limit (which is unreachable in practice), the effects of 
numerical diffusivity are unimportant. However, for any finite cell size of practical use 
(especially in 3D models), numerical diffusion does not necessarily mimic the effects of 
physical dissipation, particularly when the numerical discretization of the PDEs is performed 
employing high-order methods. It is therefore sound to assess the effects of numerical 
diffusion as a source of dissipation specifically in (GR)FFE, which should, in theory, not 
incorporate any physical dissipation.

We have performed a number of tests to quantify the amount of numerical dissipation as a 
function of resolution in our numerical algorithm. For that purpose, we have employed tests 
in 1D and 2D, but not 3D tests. This is, firstly, due to the lack of straightforward genuinely 
3D analytic solutions including resistivity in the literature. Secondly, 3D tests are computationally much 
more expensive than 1D or 2D numerical experiments. Nevertheless, we assume that the 
multidimensional nature of 2D tests presented should also shed light on the numerical dissipation 
in 3D. Numerical experience tells us that the dissipation introduced by algorithms based 
upon a directional (split) integration of the equations is only (very) weakly dependent on 
dimensionality. Backing up these assumption, we have shown here in this work that the level 
of numerical resistivity depends on the number of zones per characteristic size to be resolved 
in every dimension (compare the values of $\eta_*$ in Sect.~\ref{sec:PlasmaWaves} for 1D, 
Sect.~\ref{sec:TearingModes} for 2D), and not on the dimensionality of the problem.

In the following sections, we address the similarities and differences between numerical 
and physical resistivity in FFE. The aim of this discussion is to provide support for the 
interpretation of the numerical dissipation found in astrophysical applications of our 
code \citep{Mahlmann2019,Mahlmann2020} as a consistent model of physical dissipation. Along 
the way, we may also contribute to asses whether suitably extended FFE models (including 
physical dissipation, albeit not at the astrophysically expected levels) may be used to 
study a number of dissipative processes in relativistic astrophysics, for example, resistive solutions 
for pulsar magnetospheres \citep{Li2012}, the dissipation induced by the kink instability in 
relativistic jets \citep{Bromberg2019}, or the conversion between fast and Alfv\'{e}n 
modes \citep{Li2019}.

\subsection{Tearing modes in ideal FFE}

In physically driven TMs, the resistive effects are restricted to the so-called resistive 
layer, with width $L_{\rm R}\ll a$ for inviscid, incompressible MHD plasma \citep[e.g.,][]{Furth_etal:1963}. 
Resolving the resistive layer numerically is challenging, due to the hierarchy of scales 
$L_{\rm R}\ll a \ll L$ that must be spanned in these experiments \citep{Rembiasz2017}. 
$L_{\rm R}$ depends on resistivity and TM growth rate \citep{Rembiasz2017}:
\begin{align}
	L_{\rm R} \approx 1.48 \left(\frac{\gamma_\textsc{tm} \eta_*a^2}{k^2 v_{\rm A}^2} \right)^{1/4} = 1.48 a \frac{\left(\gamma_\textsc{tm} \eta_*\right)^{1/4}}{\left(k a\right)^{1/2}},
	\label{eq:LR}
\end{align}
where we have replaced the physical resistivity by the numerical one in account of the fact 
that the TMs in our models employing the force-free current $I^\mu_\textsc{ff}$ are induced 
by numerical resistivity. Also, we used $v_{\rm A}=1$ (in the limit of FFE) and 
highlighted the dependence of $L_{\rm R}$ on the product $(\gamma_\textsc{tm}\eta_*)^{1/4}$. 

The growth rate, $\gamma_\textsc{tm}$, and the physical resistivity, $\eta$, depend on each other 
(Eqs.~\ref{eq:gammatm_2} and \ref{eq:gammatm_3}). However, the relation between 
$\gamma_\textsc{tm}$ and the numerical resistivity $\eta_*$ is, a priori, unknown. 
This is the reason to test more than a single possible limit in Sect.~\ref{sec:TearingModes} 
(i.e. to consider models A and B). In both, $\gamma_\textsc{tm}\rightarrow 0$ and 
$\eta_*\rightarrow 0$ as the grid spacing decreases; the rate at which they approach zero 
depends on the order of the method. Hence, there is no guarantee that the product 
$\gamma_\textsc{tm}\eta_*$ in Eq.~\eqref{eq:LR} tends toward a finite (nonzero) value as the 
grid spacing is decreased (independently of the order of spatial convergence of the method). 

The width of the resistive layer in our numerical experiments can be approximately obtained 
by the half-distance between the two extrema of the component of the drift velocity associated 
with the growth of the TMs 
\citep[$V^z$ in our setup, see Fig.~\ref{fig:VelocityProfile}; cf.][Fig.~10]{Rembiasz2017}. 
We locate the position of these velocity extrema (measured at $x=0.5$) within 
$|z|\lesssim 2\Delta z$, and typically, at $z=\pm \Delta z$. In practice, these results 
imply that for every resolution we use, the numerical value $L_{\rm R}\lesssim \Delta z$ 
and, hence, there is no possibility of finding a convergent behavior. The finer the grid 
spacing, the smaller $L_{\rm R}$. A finer grid spacing yields smaller $\eta_*$, which 
translates into a shorter timescale of energy-momentum loss (see below). This prevents the 
resistive layer to fully reach the conditions under which the growth of TMs would develop 
theoretically.

In order to cross-validate these findings, we assume that the numerical TMs develop in 
regimes where the approximations that hold for model A or model B are correct. 
Plugging Eq.~\eqref{eq:gammatm_2} or Eq.~\eqref{eq:gammatm_3} into Eq.~\eqref{eq:LR}, one obtains two 
expressions for $L_{\rm R}(\gamma_\textsc{tm})$, which implicitly depend on the grid spacing. 
For the two models A and B, the right panel of Fig.~\ref{fig:TMGrowth2} shows the ratio 
$L_{\rm R}(\gamma_\textsc{tm})/\Delta z$ for different resolutions, and different methods 
of cell interface reconstruction. The obtained values $L_{\rm R}(\gamma_\textsc{tm})/\Delta z<1$ 
demonstrate that the resistive layer width is unresolved: Less than one grid zone covers the 
central region of the current layer, confirming our direct measurement of $L_{\rm R}$ 
(see above). Interestingly, the estimated value of $L_{\rm R}(\gamma_\textsc{tm})/\Delta z$ 
depends on the spatial order of accuracy of the method. The highest-order reconstruction (MP9) 
yields the smallest $\gamma_\textsc{tm}\eta_*$, while the second-order accurate MC method 
allows us to resolve the resistive layer width very marginally as 
$L_{\rm R}(\gamma_\textsc{tm})/\Delta z\gtrsim 1.5$ if more than $p=20$ grid zones span the 
current width $a$.

In the light of the previous considerations, we can now discuss our assumption of a fixed 
characteristic length, independent of resolution \eqref{eq:Lcharacteristic}. For that, we 
have also analyzed the results following more closely the procedure of \cite{Rembiasz2017}: 
We use $\mathcal{L}=L_{\rm R}$, substitute expression~\eqref{eq:LR} into Eq.~\eqref{eq:NumResistivityDx}, 
and isolate $\eta_*$ as a function of $\gamma_{\textsc{tm}}$, $\Delta z$, and the remaining 
parameters of the problem. This expression $\eta_*(\gamma_{\textsc{tm}},\Delta z)$ is then 
provided to Eq.~\eqref{eq:gammatm_1}, which yields a power-law dependence 
$\gamma_\textsc{tm}\propto \Delta z^{3r/(3+2r)}$ for model A. Analogously, for model B one 
obtains $\gamma_\textsc{tm}\propto \Delta z^{4r/(5+3r)}$. Thus, the slope of the relation in Eq.~\eqref{eq:loggammavslogdx} is related with the order of convergence $r$ through
\begin{align}
	m=-3r/(3+2r) \Leftrightarrow r &= -3 m/(3 + 2 m) \qquad \text{(model A)},\label{eq:rA}\\
	m=-4r/(5+3r) \Leftrightarrow r &= -5 m/(4 - 3 m) \qquad \text{(model B)}.\label{eq:rB}
\end{align}
Evaluating $r$ using Eqs.~\eqref{eq:rA} or~\eqref{eq:rB} yields inaccurate results, since the 
assumption that $L_{\rm R}$ is given by Eq.~\eqref{eq:LR} is only approximated (as shown above). 
The degree of inaccuracy is roughly quantified in the last column of Table~\ref{tab:tearing_table}, 
where we show the deviation of the fit parameter $m$ with respect to its analytical value, 
$m_a$, assuming that $r$ is equal to the formal order of convergence of the respective method 
(i.e., we calculate $\Delta m/m:=1-m_a/m$). The empirical order of convergence is very sensitive 
to the exact value of $m$. This sensitivity is rooted on the dependence of $L_{\rm R}$ on $\Delta z$. 
We evaluate this dependence in our results in the right panel of Fig.~\ref{fig:TMGrowth2}. 
Repeating the exercise of the previous paragraph but isolating $L_{\rm R}$, one finds a 
dependence with resolution $L_{\rm R}/\Delta z \propto \Delta z^{m_L}$, where 
\begin{align}
	m_L&=-3/(3+2r) \qquad \text{(model A)},\label{eq:mLA} \\
	m_L&=-5/(5+3r) \qquad \text{(model B)}.\label{eq:mLB}
\end{align}
Therefore, $m_L\lesssim 0$ for positive values of $r$. A direct comparison with the data in 
the right panel of Fig.~\ref{fig:TMGrowth2} suggests a small positive value of the exponent 
$m_L$, namely, $m_L\gtrsim 0$. We also highlight the fact that a value $m_L\approx 0$ would 
not allow one to reliably obtain $r$ from the slope of the linear fit in Eq.~\eqref{eq:loggammavslogdx}. 
Therefore, assuming that the physical relation of Eq.~\eqref{eq:LR} holds is not appropriate in our 
models.

Admittedly, the assumption made in Eq.~\eqref{eq:Lcharacteristic} is simplistic, but it serves 
for a double purpose. First, it states that $\mathcal{L}\ll a$. Second, it provides an order 
of magnitude estimate of the numerical resistivity. A direct inspection of Fig.~\ref{fig:TMGrowth} 
shows that $\eta_*$ is below $10^{-4}$ ($2\times 10^{-6}$) even using $p=6$ points across 
the current layer width and employing MC (MP9) reconstruction. A direct comparison with 
the results obtained by the diffusion of 1D plasma waves (Sect.~\ref{sec:PlasmaWaves}) is 
not straightforward due to the different meaning of $p$. For 1D plasma waves $p$ expresses 
the number of numerical zones per characteristic length of the problem (i.e., the wavelength 
of the waves). Here it is the number of zones per current sheet width, a scale that is (significantly) 
larger than the characteristic scale of the problem at hand (i.e., $L_{\rm R}$). Thus, the 
estimate of $\eta_*$ via TMs is less accurate than it is in the 1D case. It is a lower 
bound to the true numerical resistivity of our method as verified with the results of 
Sect.~\ref{sec:TM-physical}.

\subsection{Contrasting Numerical Resistivity in MHD and FFE}

\begin{figure}
	\centering
	\includegraphics[width=0.49\textwidth]{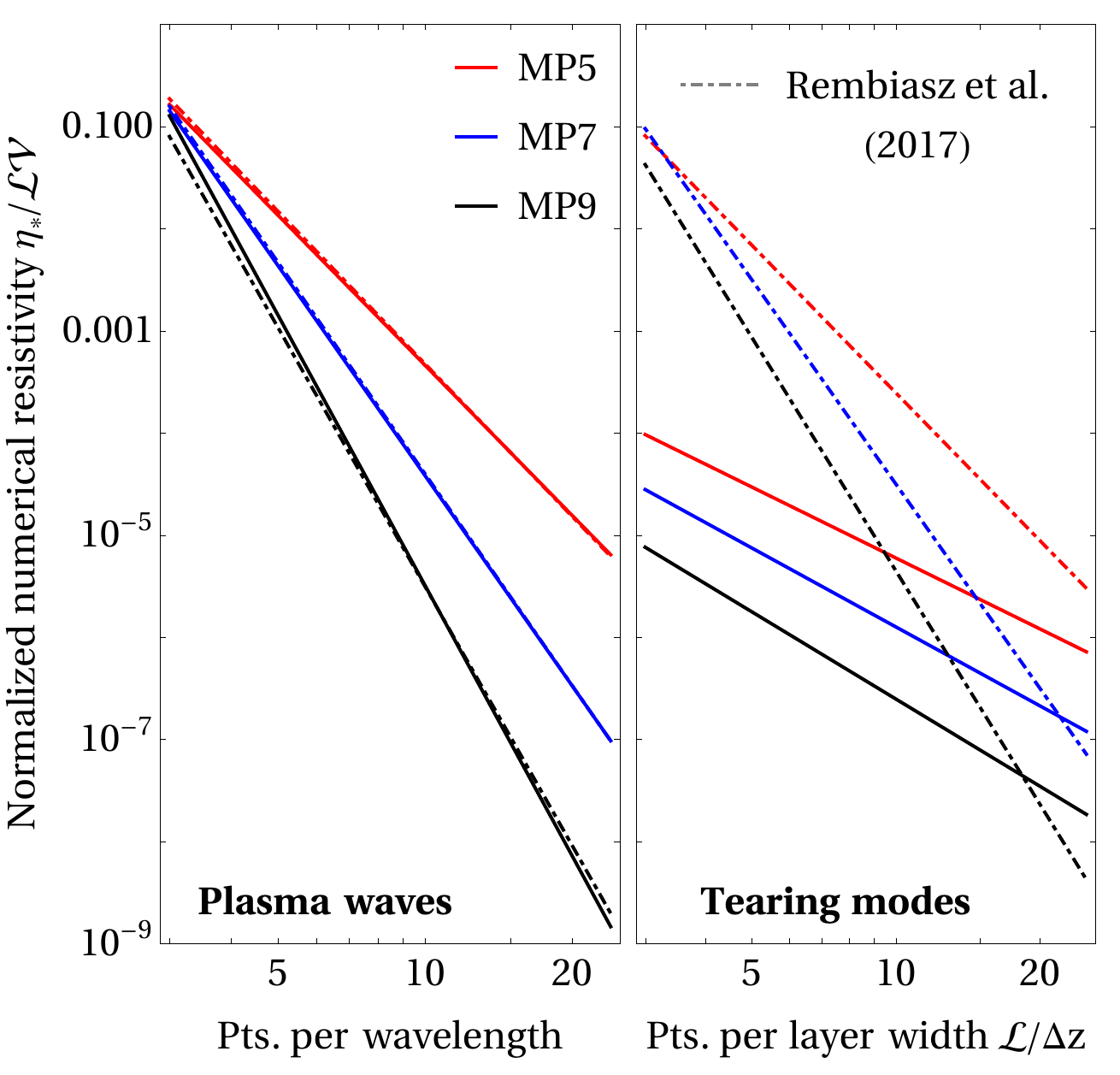}
	\caption{Comparison of normalized numerical resistivity $\eta_*/(\mathcal{L}\mathcal{V})$ ($=\mathcal{R}_m^{-1}$)
		for selected cases between the presented FFE method (solid lines) and the 3D Eulerian 
		MHD code \textsc{AENUS} \citep{Obergaulinger2008} as characterized by 
		\citet[][dot-dashed lines]{Rembiasz2017}. \textit{Left:} Comparison of the (normalized) 
		numerical resistivity as derived for (smooth) plasma waves in Sect.~\ref{sec:PlasmaWaves}. 
		\textit{Right:} Approximately contrasting the development of TMs as analyzed in 
		Sect.~\ref{sec:TearingModes}.}
	\label{fig:FFEMHDCompare}
\end{figure}

The described situation is specific to the numerical modeling of FFE:
The dissipation of energy in current sheets proceeds nearly instantaneously when the 
force-free conditions are algebraically enforced (Sect.~3.3 of Paper I). Energy-momentum 
may leak out of the system. Contrasting this, the electromagnetic energy in resistive MHD may 
be stored in other forms of energy (kinetic or thermal) as resistive processes develop within 
the system. Indeed, even when employing an ideal MHD numerical modeling, the numerical resistivity 
may find channels to convert magnetic energy-momentum into other dynamical components of the 
system \citep[e.g.,][]{Rembiasz2017}. In Fig.~\ref{fig:FFEMHDCompare} we directly compare the 
estimates of numerical resistivity determined in Sects.~\ref{sec:PlasmaWaves} and~\ref{sec:TearingModes}, 
for our FFE code, with the characterization of the 3D Eulerian MHD code \textsc{AENUS} 
\citep{Obergaulinger2008,Rembiasz2017}. The contrasting results for plasma waves and TMs lucidly illustrate a key feature of FFE methods: While smooth plasma waves are resolved 
with very competitive accuracy, comparable to the reference MHD code, the limits of ideal FFE 
are reached in resistive layers and discontinuities (e.g., current sheets). The presented 
comparison of TMs (right panel of Fig.~\ref{fig:FFEMHDCompare}) is merely an 
approximation due to the physical differences between the FFE and MHD regime. The fact that 
our FFE method shows less numerical diffusion than the MHD reference at small resolutions 
must be taken with care. Rather than an improvement of the method, this signature highlights 
the fact that, in the presence of current sheets, numerical resistivity in FFE does not 
necessarily behave as the physical one.

The physical conditions under which FFE is valid may mimic, to some extend, plasma regimes 
The enforcement of the constraints of FFE is an intrinsically nonconservative process, acting 
almost instantaneously (see Sect.~3.3 of Paper I). As a direct consequence, energy-momentum 
may leak out of the numerical domain over timescales as small as the simulation time step. 
Differently from (relativistic) MHD, the numerical resistivity of an FFE code not only depends on the discretization of the partial differential equations, but also on the degree 
by which the FFE constraints are violated during the time evolution. The amount by which FFE 
conditions are not preserved does not necessarily decrease with increasing resolution. 
Thus, the numerical resistivity effectively acts as a timescale for the numerical dissipation 
of the electromagnetic energy-momentum, where smaller values of $\eta_*$ yield shorter cooling 
timescales in the sense that ideal (perfectly conducting) FFE most efficiently dissipates 
nonideal electric fields. However, the effects of cooling are not specifically accounted 
for in the derivation of the dispersion relation of TMs (in our case Eqs.~\ref{eq:gammatm_1} 
and~\ref{eq:gammatm_2}). We may interpret that the standard TM dispersion relation 
(which strictly holds in adiabatic, incompressible MHD) is adequate for conditions 
of extremely slow cooling rather than in the limit of effective cooling that the FFE conditions 
impose. As $\eta_*$ is predominantly dictated by the order of the method, only FFE methods 
of relatively low-order may yield numerical dissipation timescales long enough that conditions 
of (sufficiently) slow cooling hold. However, this feature significantly differentiates FFE 
from (relativistic) MHD, where $\eta_*$ cannot be interpreted as a cooling timescale at all.

\subsection{The role of $\mathbf{j}_{||}$ in FFE codes}

The part of the force-free current which is along the magnetic field ($\mathbf{j}_{||}$) is 
closely related to the force free conditions: The current $I_\textsc{ff}^\mu$ conserves the 
perpendicularity of the electromagnetic fields. This can be done either by combining 
$I^\mu_\textsc{ff}$, namely, the condition $\mathcal{L}_{n}  (\mathbf{B}\cdot\mathbf{D})=0$ 
or its flat space equivalent $\partial_t\left(\mathbf{D}\cdot\mathbf{B}\right)=0$, with an 
algebraic enforcement of perpendicularity. Alternatively, the algebraic enforcement can be 
replaced by suitable Ohm's laws in the limit of low resistivity, as the one employed by 
\citet{Parfrey2017}, which we probed in Sect.~\ref{sec:Beyond_Ideal}.

As commented above, not enforcing an instantaneous fulfilment of the FFE conditions is 
equivalent to introducing a finite timescale for (resistive) dissipation of energy-momentum 
in the system of equations. If this finite timescale is long enough, the effective cooling 
may be relatively small during a sizable fraction of the linear phase of TM development. 
Hence, conditions of relatively slow (inefficient) cooling can be phenomenologically mimicked in this way. However, beyond the timescale mentioned above, energy-momentum yet 
continues to leak out of the system. This energy is not converted to other dynamical forms, 
again, differing from the physical mechanisms of energy-momentum transformation operating 
in resistive MHD.

The left panel of Fig.~\ref{fig:TMGrowth2} shows the effect of different methods for the 
discretization of $\mathbf{j}_{||}$ on the measured TM growth rates. The solid lines 
correspond to our default fourth-order accurate discretization of $\mathbf{j}_{||}$ in 
combination with different methods of cell interface reconstruction and a fixed number of 
points per current sheet width $p=6$. The dashed lines correspond to taking 
$\mathbf{j}_{||}=\mathbf{0}$ \citep[following][]{Yu2011}. Clearly, employing 
$\mathbf{j}_{||}=\mathbf{0}$ yields a significantly smaller growth rate and, hence a 
significantly smaller numerical resistivity. However, employing this procedure, the 
three-wave Riemann problem renders nonphysical results as the reduced numerical diffusivity 
of the method allows for the breakup of the fast waves into a pair of discontinuities 
(one of which should not exist; see top left panel of Fig.~\ref{fig:JparJperp}). We further probed this specific pathology in the context of nonideal FFE in 
Sect.~\ref{sec:1D_Charge_Carrying}. We attribute the inaccurate modeling of charge 
carrying current-layers in this particular case ($\mathbf{j}_{||}=\mathbf{0}$, but enforcing 
the force-free conditions algebraically) to the continuity equation of charge which we 
evolve in our method \citep{Mahlmann2020b}. Not supplying a current which is consistent 
with the plasma dynamics, namely, altering the electric field algebraically without feedback 
to the currents and charges they support, introduces errors into the numerical solution 
that will eventually lead to nonphysical behavior. For reference, we also display our 
default fourth-order discretization of $\mathbf{j}_{||}$ (dash-dotted lines). However, 
in this case, we do not enforce the algebraic fulfillment of the $\mathbf{D}\cdot\mathbf{B}=0$ 
condition. This is an extreme case (or upper bound) in terms of numerical resistivity since it yields the largest growth rate of all the models displayed in Fig.~\ref{fig:TMGrowth2}. 
The growth is so fast, that results with different cell interface reconstructions are 
indistinguishable and all the lines are located on top of each other. From these results, 
we conclude that without resorting to phenomenological approaches to introduce resistive 
dissipation in the current parallel to the magnetic field, the limits of FFE as a zeroth-order 
approximation of relativistic MHD are reached (and breached) in current sheets.

At this point, it is important to highlight that alternative expressions of the force-free 
current $I_\textsc{ff}^\mu$ (Eq.~\ref{eq:FFCurrent}) may help to increase the timescale over which electromagnetic dissipation in FFE is generated. These alternative 
expressions of the spatial current encode different forms of Ohm's law in order to derive the 
functional form of $\mathbf{j}_{||}$ 
\citep[see, e.g., Sect.~2.3 of][for a comprehensive overview and also \citealt{Lyutikov2003,Li2012,Parfrey2017}]{Alic2012}. 
Such alternative Ohm's laws relax the strict fulfillment of the FFE conditions temporarily. 
Instead, the FFE constraints are asymptotically enforced with suitable driving terms 
(see Sect.~3.3 of Paper I) or by adopting some phenomenological form of anisotropic 
electric conductivity, similar to the approach we have introduced in Sect.~\ref{sec:Beyond_Ideal}. 
There, we employ and test the isotropic resistivity model suggested in \citet{Parfrey2017}, 
which induces a simple Ohm's law of the form $\mathbf{D}_{||}\rightarrow\alpha^{-1}\eta\mathbf{J}_{||}$. 
Several other choices of suitable Ohm's laws are imaginable and found at least to some extend throughout the literature, for example, with different values along and across the magnetic field lines \citep[][]{Komissarov2004}. By employing such techniques, \citet{Komissarov_etal_2007MNRAS.374..415} 
reproduce the theoretical dispersion relation for TMs roughly, employing a second-order accurate 
method in combination with a (phenomenological) anisotropic conductivity. As such, it 
corresponds to a plasma with seemingly large resistivity ($\eta=10^{-4}-10^{-3}$), and even 
qualitatively reproduces the linear phase of collapse of stressed current sheets as compared 
with PIC simulations \citep{Lyutikov2017}. Future upgrades of our methodology will explore 
the impact of different forms of the Ohm's law in the numerical diffusivity of the algorithm 
more quantitatively; in Sect.~\ref{sec:Beyond_Ideal} we have characterized one specific 
prescription of resistive FFE in this context.

\section{Conclusions}
\label{sec:conclusions}

We have very carefully assessed the numerical resistivity of our new GRFFE code. 
This topic is of the utmost importance to interpret the dissipation of magnetic flux and 
energy in magnetically dominated astrophysical scenarios. The intrinsic numerical resistivity 
of our code depends on the number of numerical cells per characteristic size of the structure 
we aim to resolve. We have considered two types of tests. 1) The damping of 1D plasma waves. 
Such ideal (smooth) waves allow for a direct comparison of the damping rates with existing 
analytic results in the literature. 2) The growth of TMs, induced by the action of the 
numerical resistivity, in a simple 2D setup. The growth rate of the TMs depends (nonlinearly) 
on the numerical resistivity. We have performed an extensive suite of tests with different 
resolutions, spatial reconstruction methods and different forms of Ohm's law (i.e., 
different closure relations for the current as a source of the electromagnetic field and 
flux of charge conservation). With this strategy, we have assessed the signatures of numerical 
resistivity in our specific method, and FFE codes in general. We have related these 
findings to the employed numerical techniques as well as to the grid spacing employed for 
the discretization of the plasma continuum.

Resolving a characteristic length $\mathcal{L}$ with at least $\sim 5$ (10) numerical zones 
is required to reduce the numerical resistivity to values below $\eta_*\sim 10^{-2}$ 
($\eta_*\sim 10^{-4}$). In practical applications of our code to, for example, global 3D models of 
magnetar magnetospheres \citep{Mahlmann2019} we employ a typical resolution of 32 zones 
per radius of the magnetar, $R_\ast$. Hence, at scales of $1R_\ast$ our magnetosphere models 
incorporate a numerical resistivity of $\sim 10^{-5}- 10^{-8}$ (MP5 vs. MP9 reconstruction). 
In the same models, we observe that there exists dissipation of magnetic energy on structures 
with sizes as small as $\sim 0.1R_{\ast}$ \citep{Mahlmann2019}. At these scales, the numerical 
resistivity can be as large as $\eta_*\sim10^{-2}$. Similar comments are in place for global 
3D models of BH magnetospheres, where we have also used 32 cells per gravitational radius 
in the finest grid surrounding the BH \citep[but 16 zones in the extended region surrounding 
the BH][]{Mahlmann2020}. Finding dissipation on scales on the order of one gravitational radius 
in the vicinity of the BH (induced by the generation of local current sheets) means that the 
numerical resistivity of our code was $\lesssim 10^{-5}$ in these (small) regions of the 
whole computational domain. We highlight the non-uniform character of numerical resistivity 
in practical applications. For sufficiently small scales, a GRFFE code may become extremely resistive. 
This justifies the implementation of very high-order spatial reconstruction methods in GRFFE, 
even at the cost of some parallel performance loss due to the increased number of (internal) 
boundary zones that need to be synchronized. As we have shown, for resolutions at reach 
in global 3D numerical simulations (say of $5-10$ cells per length scale to be resolved) 
using a seventh-order accurate method as MP7 reduces the numerical resistivity by more than two 
orders of magnitude compared to, for example, employing a second-order accurate method 
such as MC reconstruction.

The tested resistive FFE using a simple Ohm's law \citep[Eq.~\ref{eq:FFResCurrent}, and][]{Parfrey2017} 
is able to induce the expected phenomenological behavior in a current-carrying discontinuity 
(Sect.~\ref{sec:1D_Charge_Carrying}): With increasing (physical) resistivity $\eta$, the 
current (and associated charge density) sustaining a standing Alfv\'{e}n wave becomes weaker 
and the layer diffuses (Fig.~\ref{fig:TMResGrowth}). Eventually, the solution becomes 
nonphysical and sharp transition layers emerge. On the other hand, the force-free solution 
is recovered for low values of $\eta$. In the case of TMs, the considered Ohm's law converges 
toward growth rates comparable to the ones expected by the dynamics driven by a physical 
resistivity when endowed with $\eta$ above the threshold of numerical resistivity.

We are, in conclusion, able to give three answers to the initially posed questions: 
a) With an accurate modeling of the current $\mathbf{j}_\parallel$ 
(including a sufficiently high-order discretization of it), our finite volume ideal 
FFE code resolves smooth current-carrying plasma waves with close-to-exact accuracy. 
b) In force-free current sheets a true limit of ideal FFE is reached; resistive layers cannot 
be resolved on distances on the order of the grid-scale. 
c) When reducing the numerical resistivity by increasing resolution or the numerical accuracy 
(increasing the order of the spatial reconstruction), growth rates of the considered dynamical 
instability (TMs) in FFE are not fully comparable with their counterpart in numerical 
experiments, for example, in resistive MHD. The estimated order of accuracy of the FFE method is reduced 
in such regions. The action of numerical resistivity in the limit of validity of FFE does 
not exactly match the action of physical resistivity. In other words, numerical resistivity 
does not necessarily mimic physical resistivity in tearing-unstable current sheets. Alternative 
Ohm's laws that incorporate a physical resistivity are promising candidates for a correct 
physical modeling of genuinely resistive effects in FFE. Thus, they may be used as an algorithmic 
strategy to bridge between plasma regimes in hybrid simulation tools combining, for example, FFE with 
particle in cell simulations \citep{Parfrey2019}. The numerical techniques presented in this 
series of publications are able to resolve the global dynamics of force-free plasma with a great 
consistency of its physical constituents and achieves competitive accuracy.

\section{Acknowledgments}

We appreciate the helpful comments and perspectives contributed by the anonymous referee. We thank B. Ripperda and A. Philippov for their perspectives on plasma (codes). We acknowledge the fruitful discussions with T. Rembiasz, as well as his critical revision of the results related to the characterization of the numerical diffusivity of our code. JM acknowledges a Ph.D. grant of the \textit{Studienstiftung des Deutschen Volkes}. 
We acknowledge the support from the grants AYA2015-66899-C2-1-P, PGC2018-095984-B-I00, 
PROMETEO-II-2014-069, and PROMETEU/2019/071. We acknowledge the partial support of the 
PHAROS COST Action CA16214 and GWverse COST Action CA16104. 
PCD acknowledges the Ramon y Cajal funding (RYC-2015-19074) supporting his research. 
VM is supported by the Exascale Computing Project (17-SC-20-SC), a collaborative effort 
of the U.S. Department of Energy (DOE) Office of Science and the National Nuclear Security 
Administration. Work at Oak Ridge National Laboratory is supported under contract DE-AC05-00OR22725 
with the U.S. Department of Energy. VM also acknowledges partial support from the National 
Science Foundation (NSF) from Grant Nos.\ OAC-1550436, AST-1516150, PHY-1607520, 
PHY-1305730, PHY-1707946, and PHY-1726215 to Rochester Institute of Technology (RIT). 
The shown numerical simulations have been conducted on infrastructure of the 
\textit{Red Espa\~{n}ola de Supercomputación} (AECT-2020-1-0014) as well as of the 
\textit{University of Valencia} Tirant and LluisVives supercomputers. 

\bibliographystyle{aa}
\bibliography{literature.bib}

\begin{thebibliography}{41}
\expandafter\ifx\csname natexlab\endcsname\relax\def\natexlab#1{#1}\fi

\bibitem[{{Alic} {et~al.}(2012){Alic}, {Moesta}, {Rezzolla}, {Zanotti}, \&
  {Jaramillo}}]{Alic2012}
{Alic}, D., {Moesta}, P., {Rezzolla}, L., {Zanotti}, O., \& {Jaramillo}, J.~L.
  2012, \apj, 754, 36

\bibitem[{Ball {et~al.}(2019)Ball, Sironi, \& {\"O}zel}]{Ball2019}
Ball, D., Sironi, L., \& {\"O}zel, F. 2019, \apj, 884, 57

\bibitem[{Blandford \& Znajek(1977)}]{Blandford1977}
Blandford, R.~D. \& Znajek, R.~L. 1977, \mnras, 179, 433

\bibitem[{Bromberg {et~al.}(2019)Bromberg, Singh, Davelaar, \&
  Philippov}]{Bromberg2019}
Bromberg, O., Singh, C.~B., Davelaar, J., \& Philippov, A.~A. 2019, \apj, 884,
  39

\bibitem[{{Campos}(1999)}]{Campos1999PhPl....6...57}
{Campos}, L.~M.~B.~C. 1999, Physics of Plasmas, 6, 57

\bibitem[{Carrasco {et~al.}(2019)Carrasco, Vigan{\`o}, Palenzuela, \&
  Pons}]{Carrasco_etal_2019_10.1093/mnrasl/slz016}
Carrasco, F., Vigan{\`o}, D., Palenzuela, C., \& Pons, J.~A. 2019, Monthly
  Notices of the Royal Astronomical Society: Letters, 484, L124

\bibitem[{{Del Zanna} {et~al.}(2016){Del Zanna}, {Papini}, {Landi}, {Bugli}, \&
  {Bucciantini}}]{DelZanna2016}
{Del Zanna}, L., {Papini}, E., {Landi}, S., {Bugli}, M., \& {Bucciantini}, N.
  2016, \mnras, 460, 3753

\bibitem[{{Furth} {et~al.}(1963){Furth}, {Killeen}, \&
  {Rosenbluth}}]{Furth_etal:1963}
{Furth}, H.~P., {Killeen}, J., \& {Rosenbluth}, M.~N. 1963, Physics of Fluids,
  6, 459

\bibitem[{{Gruzinov}(2007)}]{Gruzinov2007}
{Gruzinov}, A. 2007, arXiv e-prints, arXiv:0710.1875

\bibitem[{Guo {et~al.}(2019)Guo, Li, Daughton, Kilian, Li, Liu, Yan, \&
  Ma}]{Guo2019}
Guo, F., Li, X., Daughton, W., {et~al.} 2019, \apjl, 879, L23

\bibitem[{Harra \& Mason(2004)}]{Harra2004}
Harra, L. \& Mason, K. 2004, Space science (Imperial College Press)

\bibitem[{Kilian {et~al.}(2020)Kilian, Li, Guo, \& Li}]{Kilian2020}
Kilian, P., Li, X., Guo, F., \& Li, H. 2020, arXiv e-prints, arXiv:2001.02732

\bibitem[{Komissarov(2004)}]{Komissarov2004}
Komissarov, S.~S. 2004, \mnras, 350, 427

\bibitem[{{Komissarov}(2011)}]{Komissarov2011}
{Komissarov}, S.~S. 2011, \mnras, 418, L94

\bibitem[{{Komissarov} {et~al.}(2007){Komissarov}, {Barkov}, \&
  {Lyutikov}}]{Komissarov_etal_2007MNRAS.374..415}
{Komissarov}, S.~S., {Barkov}, M., \& {Lyutikov}, M. 2007, \mnras, 374, 415

\bibitem[{{Li} {et~al.}(2012){Li}, {Spitkovsky}, \& {Tchekhovskoy}}]{Li2012}
{Li}, J., {Spitkovsky}, A., \& {Tchekhovskoy}, A. 2012, \apj, 746, 60

\bibitem[{Li {et~al.}(2019)Li, Zrake, \& Beloborodov}]{Li2019}
Li, X., Zrake, J., \& Beloborodov, A.~M. 2019, \apj, 881, 13

\bibitem[{{Low}(1973)}]{Low_1973ApJ...181..209}
{Low}, B.~C. 1973, \apj, 181, 209

\bibitem[{Lyutikov(2003)}]{Lyutikov2003}
Lyutikov, M. 2003, \mnras, 346, 540

\bibitem[{{Lyutikov} {et~al.}(2018){Lyutikov}, {Komissarov}, {Sironi}, \&
  {Porth}}]{Lyutikov_2018JPlPh..84b6301}
{Lyutikov}, M., {Komissarov}, S., {Sironi}, L., \& {Porth}, O. 2018, Journal of
  Plasma Physics, 84, 635840201

\bibitem[{Lyutikov {et~al.}(2017)Lyutikov, Sironi, Komissarov, \&
  Porth}]{Lyutikov2017}
Lyutikov, M., Sironi, L., Komissarov, S.~S., \& Porth, O. 2017, Journal of
  Plasma Physics, 83

\bibitem[{Mahlmann {et~al.}(2019)Mahlmann, Akg{\"u}n, Pons, Aloy, \&
  Cerd{\'a}-Dur{\'a}n}]{Mahlmann2019}
Mahlmann, J.~F., Akg{\"u}n, T., Pons, J.~A., Aloy, M.~A., \&
  Cerd{\'a}-Dur{\'a}n, P. 2019, \mnras, 490, 4858

\bibitem[{Mahlmann {et~al.}(2020{\natexlab{a}})Mahlmann, Aloy, Mewes, \&
  Cerd{\'a}-Dur{\'a}n}]{Mahlmann2020b}
Mahlmann, J.~F., Aloy, M.~A., Mewes, V., \& Cerd{\'a}-Dur{\'a}n, P.
  2020{\natexlab{a}}, arXiv e-prints, arXiv:2007.06580

\bibitem[{Mahlmann {et~al.}(2020{\natexlab{b}})Mahlmann, Levinson, \&
  Aloy}]{Mahlmann2020}
Mahlmann, J.~F., Levinson, A., \& Aloy, M.~A. 2020{\natexlab{b}}, \mnras

\bibitem[{{Miranda-Aranguren} {et~al.}(2018){Miranda-Aranguren}, {Aloy}, \&
  {Rembiasz}}]{Miranda-Aranguren2018}
{Miranda-Aranguren}, S., {Aloy}, M.~A., \& {Rembiasz}, T. 2018, \mnras, 476,
  3837

\bibitem[{{Miranda-Aranguren} {et~al.}(2014){Miranda-Aranguren}, {Aloy}, \&
  {Aloy}}]{Miranda-Aranguren2014IAUS..302...64A}
{Miranda-Aranguren}, S.~M., {Aloy}, M.~A., \& {Aloy}, C. 2014, in IAU
  Symposium, Vol. 302, Magnetic Fields throughout Stellar Evolution, ed.
  P.~{Petit}, M.~{Jardine}, \& H.~C. {Spruit}, 64--65

\bibitem[{Nathanail {et~al.}(2020)Nathanail, Fromm, Porth, Olivares, Younsi,
  Mizuno, \& Rezzolla}]{nathanail2020}
Nathanail, A., Fromm, C.~M., Porth, O., {et~al.} 2020, arXiv e-prints,
  arXiv:2002.01777

\bibitem[{Obergaulinger(2008)}]{Obergaulinger2008}
Obergaulinger, M. 2008, PhD thesis, Max-Planck-Institut f{\"u}r Astrophysik,
  Garching bei M{\"u}nchen

\bibitem[{{Obergaulinger} \& {Aloy}(2020)}]{Obergaulinger_2020arXiv200101927}
{Obergaulinger}, M. \& {Aloy}, M.-{\'A}. 2020, arXiv e-prints, arXiv:2001.01927

\bibitem[{{Palenzuela} {et~al.}(2009){Palenzuela}, {Lehner}, {Reula}, \&
  {Rezzolla}}]{Palenzuela2009}
{Palenzuela}, C., {Lehner}, L., {Reula}, O., \& {Rezzolla}, L. 2009, \mnras,
  394, 1727

\bibitem[{Parfrey {et~al.}(2013)Parfrey, Beloborodov, \& Hui}]{Parfrey2013}
Parfrey, K., Beloborodov, A.~M., \& Hui, L. 2013, \apj, 774, 92

\bibitem[{{Parfrey} {et~al.}(2015){Parfrey}, {Giannios}, \&
  {Beloborodov}}]{Parfrey2015}
{Parfrey}, K., {Giannios}, D., \& {Beloborodov}, A.~M. 2015, \mnras, 446, L61

\bibitem[{{Parfrey} {et~al.}(2019){Parfrey}, {Philippov}, \&
  {Cerutti}}]{Parfrey2019}
{Parfrey}, K., {Philippov}, A., \& {Cerutti}, B. 2019, \prl, 122, 035101

\bibitem[{{Parfrey} {et~al.}(2017){Parfrey}, {Spitkovsky}, \&
  {Beloborodov}}]{Parfrey2017}
{Parfrey}, K., {Spitkovsky}, A., \& {Beloborodov}, A.~M. 2017, \mnras, 469,
  3656

\bibitem[{Petropoulou {et~al.}(2019)Petropoulou, Sironi, Spitkovsky, \&
  Giannios}]{Petropoulou2019}
Petropoulou, M., Sironi, L., Spitkovsky, A., \& Giannios, D. 2019, \apj, 880,
  37

\bibitem[{Punsly(2003)}]{Punsly2003}
Punsly, B. 2003, \apj, 583, 842

\bibitem[{{Rembiasz} {et~al.}(2017){Rembiasz}, {Obergaulinger},
  {Cerd{\'a}-Dur{\'a}n}, {Aloy}, \& {M{\"u}ller}}]{Rembiasz2017}
{Rembiasz}, T., {Obergaulinger}, M., {Cerd{\'a}-Dur{\'a}n}, P., {Aloy},
  M.-{\'A}., \& {M{\"u}ller}, E. 2017, \apjs, 230, 18

\bibitem[{Ripperda {et~al.}(2020)Ripperda, Bacchini, \&
  Philippov}]{Ripperda2020}
Ripperda, B., Bacchini, F., \& Philippov, A. 2020, arXiv e-prints,
  arXiv:2003.04330

\bibitem[{{Suresh} \& {Huynh}(1997)}]{Suresh1997}
{Suresh}, A. \& {Huynh}, H.~T. 1997, \jcop, 136, 83

\bibitem[{{van Leer}(1977)}]{vanLeer1977}
{van Leer}, B. 1977, Journal of Computational Physics, 23, 276

\bibitem[{Yu(2011)}]{Yu2011}
Yu, C. 2011, \mnras, 411, 2461

\end{thebibliography}

\end{document}